\documentclass{aa}

\usepackage{natbib}
\usepackage{txfonts}
\usepackage[dvips]{graphicx}
\usepackage{subfigure}
\usepackage{color}
\usepackage{bm}

\bibpunct{(}{)}{;}{a}{}{,}

\begin{document}

\def\kms{km\,s$^{-1}$}
\def\vradp{$\upsilon_{\rm rad,1}$}
\def\vrads{$\upsilon_{\rm rad,2}$}
\def\Msun{${\rm M}_{\sun}$}
\def\Rsun{${\rm R}_{\sun}$}
\def\logg{$\log{g}$}
\def\loggp{$\log{g}_1$}
\def\loggs{$\log{g}_2$}
\def\teff{$T_{\rm eff}$}
\def\teffp{$T_{\rm eff,1}$}
\def\teffs{$T_{\rm eff,2}$}
\def\vga{$\upsilon_\gamma$}
\def\xld{$x_{\rm LD}$}
\def\xldp{$x_{\rm LD,1}$}
\def\xlds{$x_{\rm LD,2}$}
\def\yld{$y_{\rm LD}$}
\def\yldp{$y_{\rm LD,1}$}
\def\ylds{$y_{\rm LD,2}$}
\def\Porb{$P_{\rm orb}$}
\def\Forb{$F_{\rm orb}$}
\def\Protp{$P_{\rm rot,1}$}
\def\Prots{$P_{\rm rot,2}$}
\def\logL{$\log{(L/L_{\sun})}$}
\def\logLp{$\log{(L_1/L_{\sun})}$}
\def\logLs{$\log{(L_2/L_{\sun})}$}
\def\vsini{$\upsilon\sin{i}$}
\def\vsinip{$\upsilon_1\sin{i}$}
\def\vsinis{$\upsilon_2\sin{i}$}
\def\vmicro{$\upsilon_{\rm micro}$}
\def\vmicrop{$\upsilon_{\rm micro,1}$}

\titlerunning{CoRoT\,105906206: an eclipsing binary with a $\delta$ Scuti
type pulsator}
\authorrunning{R. da Silva et al.}

\title{CoRoT\,105906206: a short-period and totally eclipsing binary \\
with a $\delta$ Scuti type pulsator
\thanks{Based on the photometry collected by the CoRoT satellite and on
spectroscopy obtained with the Sandiford spectrograph attached at the 2.1-m
telescope at McDonald Observatory (Texas, USA) and the FEROS spectrograph
mounted on the ESO 2.2-m telescope at ESO (La Silla, Chile). The CoRoT space
mission was developed and is operated by the French space agency CNES, with
participation of ESA's RSSD and Science Programs, Austria, Belgium, Brazil,
Germany and Spain.}}

\author{R. da Silva\inst{1} \and C. Maceroni\inst{1} \and
D. Gandolfi\inst{2,3,4} \and H. Lehmann\inst{3} \and A.P. Hatzes\inst{4}}

\offprints{R. da Silva,\\
\email{ronaldo.dasilva@oa-roma.inaf.it}}

\institute{
INAF, Osservatorio Astronomico di Roma, Via Frascati 33, 00040, Monte Porzio
Catone, Italy
\and
INAF, Osservatorio Astrofisico di Catania, Via S. Sofia 78, 95123, Catania,
Italy
\and
Research and Scientific Support Department, ESA/ESTEC, PO Box 299, 2200 AG,
Noordwijk, The Netherlands
\and
Th\"uringer Landessternwarte Tautenburg, Sternwarte 5, D-07778 Tautenburg,
Germany
}

\date{Received / accepted}

%
%
\abstract{}
{Eclipsing binary systems with pulsating components allow the determination
of several physical parameters of the stars, such as mass and radius, that,
when combined with the pulsation properties, can be used to constrain the
modeling of stellar interiors and evolution. Hereby, we present the results
of the study of CoRoT\,105906206, an eclipsing binary system with a
pulsating component located in the CoRoT LRc02 field.}
{The analysis of the CoRoT light curve was complemented by high-resolution
spectra from the Sandiford at McDonald Observatory and FEROS at ESO
spectrographs, which revealed a double-lined spectroscopic binary. We used
an iterative procedure to separate the pulsation-induced photometric
variations from the eclipse signals. First, a Fourier analysis was used to
identify the significant frequencies and amplitudes due to pulsations.
Second, after removing the contribution of the pulsations from the light
curve we applied the PIKAIA genetic-algorithm approach to derive the best
parameters that describe the orbital properties of the system.}
{The light curve cleaned for pulsations contains the partial eclipse of the
primary and the total eclipse of the secondary. The system has an orbital
period of about 3.694~days and is formed by a primary star with mass
$M_1$ = 2.25 $\pm$ 0.04~\Msun, radius $R_1$ = 4.24 $\pm$ 0.02~\Rsun, and
effective temperature \teffp\ = 6750 $\pm$ 150~K, and a secondary with
$M_2$ = 1.29 $\pm$ 0.03~\Msun, $R_2$ = 1.34 $\pm$ 0.01~\Rsun, and
\teffs\ = 6152 $\pm$ 162~K. The best solution for the parameters was
obtained by taking into account the asymmetric modulation observed in the
light curve, known as the O'Connell effect, presumably caused by Doppler
beaming. The analysis of the Fourier spectrum revealed that the primary
component has p-mode pulsations in the range 5-13~d$^{-1}$, which are
typical of $\delta$ Scuti type stars.}
{}


\maketitle

%
%
\section{Introduction}
\label{intro}

The study of eclipsing binary systems has gained a new perspective since the
beginning of the CoRoT space mission \citep{Baglinetal2006}. The superb
photometry achievable from space, combined with ground-based spectroscopy,
allows a precise and independent determination of mass and radius of the
components, among other parameters. In particular, by studying pulsating
stars in eclipsing binaries, such as the Classical $\gamma$~Dor and
$\delta$~Sct type variables, one takes advantage of this parameter
determination for the asteroseismic modeling of stellar structure and
evolution. The CoRoT observations unveiled several targets suitable for this
kind of research.

$\delta$~Sct type variables are stars located in the classical instability
strip on the H-R diagram with effective temperatures in the range
6300~$<$~\teff~$<$~9000~K, luminosities 0.6~$<$~\logL~$<$~2.0, and masses
between 1.5 and 2.5~\Msun. Their evolutionary stages range from pre-main
sequence to just evolved off the main sequence (about 2~mag above the ZAMS).
They exhibit radial and/or non-radial pulsations, with low-order gravity (g)
and/or pressure (p) modes with pulsation periods ranging from $\sim$15~min
to $\sim$8~h \citep[see e.g.][and references therein]{RodriguezBreger2001,
Buzasietal2005,Uytterhoevenetal2011}.

Hence, $\delta$~Sct type stars are an interesting class of objects since
they lie in the transition region between stars having a convective
($M < 2$~\Msun) or a radiative ($M > 2$~\Msun) envelope. Their masses are in
a range where stars are developing a convective zone so they are useful for
a better understanding of the mechanisms responsible for driving the
pulsations. In this work we describe the analysis of CoRoT\,105906206, an
eclipsing binary system showing properties typical of $\delta$~Sct type
variables.

Sections~\ref{phot} and \ref{spec} present the details of the photometric
and the spectroscopic observations, respectively. Section~\ref{analysis}
describes how we derive the parameters of the system through the analysis
of the light and radial-velocity curves, and Sect.~\ref{phy_prop} provides
additional physical properties. The resulting pulsation frequencies are
discussed in Sect.~\ref{puls_prop}, and our final remarks and conclusions
are in Sect.~\ref{remarks}.

%
%
\section{CoRoT photometry}
\label{phot}

CoRoT\,105906206 was observed during the second "Long Run" in the direction
of the Galactic centre (LRc02), which lasted about 161~days. According to
the information in the ExoDat database \citep{Deleuiletal2009}, this
relatively bright target ($V = 11.784 \pm 0.044$~mag) has a small level of
contamination ($L_0 = 0.0077$, where $L_0$ is the ratio between contaminant
and total fluxes) and thus was not removed from the observed flux. We used
in our analysis a white light curve, which is the sum of the three chromatic
light curves from the CoRoT three-color photometry \citep{Auvergneetal2009}.
These colors (red, green, and blue channels) do not represent any standard
photometric system. They were implemented to help with the identification of
false alarms, which can mimic the transit of a planet in front of the main
target.

\begin{figure}
\centering
\resizebox{\hsize}{!}{\includegraphics[angle=-90]{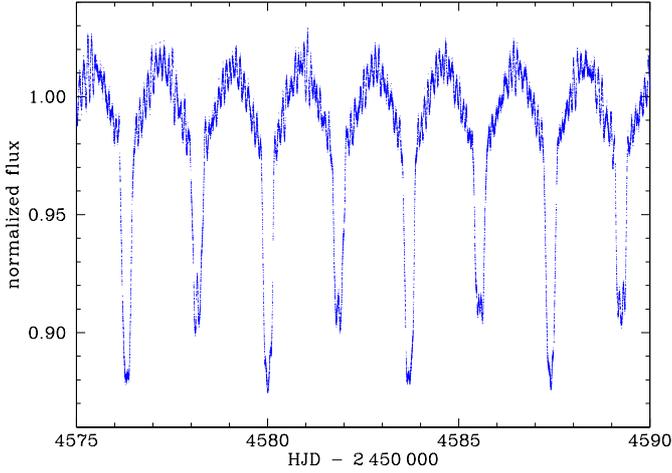}}
\caption{Portion of the CoRoT\,105906206 white light curve normalized to the
         mean value. The full data extends over 144~days.}
\label{lc_obs}
\end{figure}

The original time series, containing about 388\,000 points, was detrended to
remove long-term variations, and was cleaned from outliers using a
sigma-clipping algorithm. Points flagged according to the description of
\citet{Gruberbauer2008} were also excluded. Only a few measurements
(corresponding to about 0.5~days) were made in the long-integration mode
(512~s) and they were not used in our analysis. The resulting light curve
contains about 341\,000 points collected in the short-integration mode
(32~s), from HJD = 2\,454\,573 to 2\,454\,717~days. A portion of the
processed light curve is shown in Fig.\ref{lc_obs}.

%
%
\section{High-resolution spectroscopy}
\label{spec}

We collected high-resolution spectra of CoRoT 105906206 using two
instruments:\, the Sandiford Echelle Spectrograph \citep{McCarthyetal1993}
attached at the Cassegrain focus of the 2.1-m telescope of McDonald
Observatory (Texas, USA), and the fiber-fed FEROS Echelle Spectrograph
\citep{Kaufer1999} mounted on the MPG/ESO 2.2-m telescope at La Silla
Observatory (Chile).

Six Sandiford spectra were taken over 6 consecutive nights in May 2011 under
fairly good sky conditions, with seeing typically varying between 1.0 and
2.0~\arcsec. We set the grating angle to cover the wavelength range
5000--6000~\AA\ and used the 1.0~\arcsec\ wide slit, which yields a
resolution of $R \sim 47\,000$. We adopted an exposure time of
1200--1800~sec and traced the radial velocity drift of the instrument by
acquiring long-exposed ($T_{\rm exp}$ = 30~sec) ThAr spectra right before
and after each epoch of observation. 

Nineteen additional spectra were acquired in June-July 2011 with FEROS,
which provides a resolution of $R \sim 48\,000$ and covers the wavelength
range $\sim$3500--9200~\AA. The sky conditions were photometric throughout
the whole observation run, with seeing between 0.6 and 1.0~\arcsec. Exposure
times ranged between 1200 and 1800~seconds.


\begin{figure}
\centering
\resizebox{\hsize}{!}{\includegraphics[angle=-90]{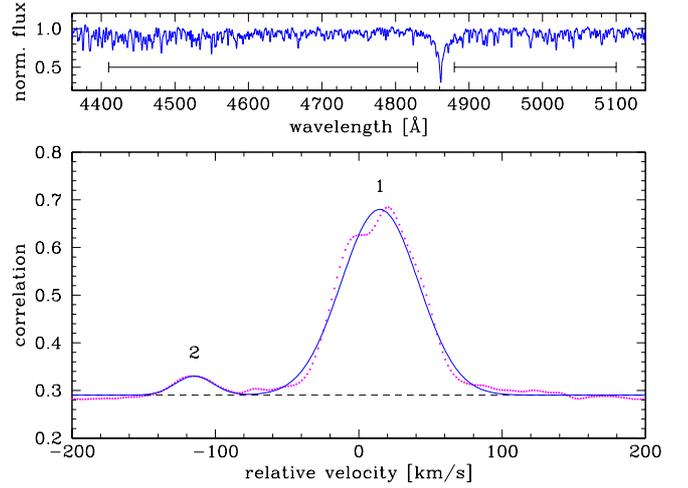}}
\caption{Example of cross-correlation used to derive the radial velocities
         of the CoRoT\,105906206 components.
         {\it Top panel:} a portion of FEROS spectra showing the regions
	 used (delimited by horizontal bars). 
	 {\it Bottom panel:} cross-correlation peaks of the primary (1) and
	 secondary (2) stars (dots) fitted by a double-Gaussian function
	 (solid line).}
\label{ccf_ex}
\end{figure}

The spectra were reduced using IRAF\footnote{{\it Image Reduction and
Analysis Facility}, distributed by the National Optical Astronomy
Observatories (NOAO), USA.} standard routines and the FEROS automatic
pipeline for order identification and extraction, background subtraction,
flat-field correction and wavelength calibration. The radial velocities of
the two components, together with their uncertainties, were derived with
the {\it fxcor} task of IRAF by cross-correlating each spectrum with a
reference template. The standard stars \object{HD\,168009}
\citep{Udryetal1999} and \object{HD\,102870} \citep{Nideveretal2002}, both
observed with the same instrumental set-up, were used as template for
Sandiford and FEROS spectra, respectively. Figure~\ref{ccf_ex} shows the
cross-correlation function (CCF) peaks fitted by a double Gaussian curve.
The uncertainty in each radial-velocity measurement, and whether one or a
double Gaussian is fitted, depend on the separation of the peaks. 

In the beginning of our study, we made a visual comparison of our spectra
with the Digital Spectral Classification Atlas of
R. O. Gray\footnote{http://ned.ipac.caltech.edu/level5/Gray/frames.html},
which suggested a spectral type F3 for the primary star, corresponding to an
effective temperature of about 6800~K. This is the value we used in
this work, though it was afterward improved by the spectroscopic analysis
of the primary decomposed spectrum, obtained through the disentangling
method (see Sect.~\ref{disent}). At any rate, both determinations are in
very good agreement with each other. Concerning the secondary companion, it
gives only a small contribution to the total flux, as can be seen in
Fig.~\ref{ccf_ex}.

\begin{figure}
\centering
\resizebox{\hsize}{!}{\includegraphics[angle=-90]{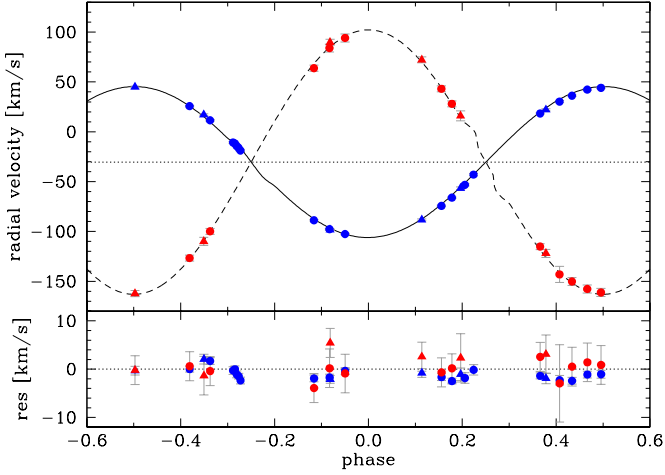}}
\caption{Phase-folded radial velocities of CoRoT 105906206 (top panel) and
         residuals of the best-fit models (bottom panel). Filled circles and
	 triangles are FEROS and Sandiford measurements, respectively.
	 Uncertainties are shown in light gray. The best Keplerian plus R-M
	 models, obtained with PHOEBE, are shown as continuous and dashed
	 lines respectively for primary and secondary components.}
\label{rv_fig}
\end{figure}

The derived radial velocities and the estimated uncertainties of the two
stars of CoRoT\,105906206 are listed in Table~\ref{rv_tab} and plotted in
Fig.~\ref{rv_fig}, together with the best-fit models computed with the
PHOEBE code \citep[PHysics Of Eclipsing BinariEs,][]{PrsaZwitter2005} for
both primary and secondary components. Besides Keplerian motion, the models
also consider the Rossiter-McLaughlin (R-M) effect.

\subsection{Spectral decomposition}
\label{disent}

The VO-version\footnote{http://vokorel.asu.cas.cz/app/} of the KOREL program
\citep{Hadrava2004,Hadrava2009} was used to decompose the FEROS spectra into
the individual stellar components. The method of spectral disentangling
was first introduced by \citet{SimonSturm1994} and later on reformulated by
\citet{Hadrava2004} in the Fourier space.

The quality of the spectral disentangling
with KOREL strongly depends on the quality of the derived line shifts of
the components (radial velocities). To determine these shifts as precise as
possible, we excluded the broad Balmer lines from this step of the analysis.
From visual inspection, we could only find one line of the secondary
component in the spectra, \ion{Mg}{ii} at 5183~\AA. A first KOREL solution
was computed on a small interval around this wavelength and then the region
was extended to the range 4915 to 5500~\AA, also allowing for variable line
strengths in the program. Any further extension in wavelength did not
improve the accuracy of the results. Due to the low frequency problem,
typical of Fourier transform-based disentangling programs like KOREL
\citep[see e.g.][]{Hensbergeetal2008}, slight undulations occurred in the
computed continua of the decomposed spectra. Therefore, we applied KOREL to
overlapping 20~nm wide bins, where we fixed the orbital parameters and time
variable line strengths to the values obtained before. Only in the case of
broad Balmer lines did we chose wider bins. Each piece of decomposed
spectrum, which covers in total a wavelength range from 4410 to 6860~\AA,
was corrected for slight continuum undulations using spline fits. Finally,
all bins were merged using weighting ramps for the overlapping parts.

KOREL delivers the decomposed spectra of the components normalized to the
common continuum of both stars. A renormalization to the individual continua
of the components can only be done by assuming a value for their flux ratio.
Since this value is very low, the line depths of the secondary are very
sensitive to the flux ratio. This, in addition to the very low
signal-to-noise (S/N) of the decomposed secondary spectrum, prevented us
from its analysis. In the case of the primary component, small deviations
from the true flux ratio cause only a second order effect.

\begin{table}
\centering
\caption[]{Radial velocities of CoRoT\,105906206.}
\label{rv_tab}
\begin{tabular}{l c r@{}l r@{}l}
\hline\hline\noalign{\smallskip}
BJD $-$ 2\,450\,000 & \parbox[c]{1.0cm}{\centering S/N {\tiny 5500~\AA}} &
\multicolumn{2}{c}{\parbox[c]{1.0cm}{\centering \vradp\ {\tiny [\kms]}}} &
\multicolumn{2}{c}{\parbox[c]{1.0cm}{\centering \vrads\ {\tiny [\kms]}}} \\
\noalign{\smallskip}\hline\noalign{\smallskip}
\multicolumn{6}{c}{\centering Sandiford spectrograph} \\
\hline\noalign{\smallskip}
5700.80506 & 40 &  $-$88&.3 $\pm$ 0.9 &     72&.2 $\pm$ 3.0 \\
5701.78240 & 17 &     22&.1 $\pm$ 1.1 & $-$122&.0 $\pm$ 4.0 \\
5702.78403 & 44 &     17&.1 $\pm$ 1.0 & $-$110&.0 $\pm$ 4.0 \\
5703.77800 & 32 &  $-$98&.4 $\pm$ 0.8 &     89&.8 $\pm$ 3.0 \\
5704.80512 & 37 &  $-$56&.6 $\pm$ 1.3 &     16&.1 $\pm$ 5.0 \\
5705.93661 & 37 &     44&.9 $\pm$ 0.9 & $-$162&.3 $\pm$ 3.0 \\
\hline\noalign{\smallskip}
\multicolumn{6}{c}{\centering FEROS spectrograph} \\
\hline\noalign{\smallskip}
5738.68288 & 65 &     18&.4 $\pm$ 0.6 & $-$115&.2 $\pm$ 3.0 \\
5738.83568 & 25 &     30&.3 $\pm$ 1.0 & $-$143&.1 $\pm$ 8.0 \\
5739.61752 & 72 &     25&.7 $\pm$ 0.6 & $-$126&.7 $\pm$ 3.0 \\
5739.77880 & 80 &     11&.5 $\pm$ 0.7 &  $-$99&.9 $\pm$ 3.0 \\
5740.59666 & 52 &  $-$88&.8 $\pm$ 0.7 &     63&.8 $\pm$ 3.0 \\
5740.71907 & 69 &  $-$97&.7 $\pm$ 0.7 &     84&.0 $\pm$ 4.0 \\
5740.84153 & 64 & $-$102&.6 $\pm$ 0.7 &     94&.1 $\pm$ 4.0 \\
5741.60028 & 52 &  $-$74&.4 $\pm$ 0.7 &     43&.1 $\pm$ 3.0 \\
5741.68255 & 60 &  $-$66&.1 $\pm$ 0.6 &     28&.0 $\pm$ 3.0 \\
5741.78475 & 52 &  $-$53&.2 $\pm$ 1.1 &       & --	    \\
5741.85308 & 60 &  $-$43&.0 $\pm$ 1.1 &       & --	    \\
5742.62804 & 71 &     36&.2 $\pm$ 0.7 & $-$150&.2 $\pm$ 4.0 \\
5742.74851 & 68 &     42&.3 $\pm$ 0.7 & $-$157&.7 $\pm$ 4.0 \\
5742.85620 & 66 &     44&.1 $\pm$ 0.8 & $-$161&.2 $\pm$ 4.0 \\
5743.65426 & 42 &  $-$10&.7 $\pm$ 0.8 &       & --	    \\
5743.66877 & 47 &  $-$11&.8 $\pm$ 0.7 &       & --	    \\
5743.68329 & 37 &  $-$14&.3 $\pm$ 0.7 &       & --	    \\
5743.69780 & 52 &  $-$16&.2 $\pm$ 0.6 &       & --	    \\
5743.71232 & 39 &  $-$18&.8 $\pm$ 0.7 &       & --	    \\
\hline
\end{tabular}
\tablefoot{The second column gives the S/N ratio per pixel at 5500~\AA.}
\end{table}

%
\begin{figure*}
\centering
\begin{minipage}[b]{0.49\textwidth}
\flushright
(a)\hspace{0.5cm}\,\\[-4cm]
\centering
\resizebox{\hsize}{!}{\includegraphics[angle=-90]{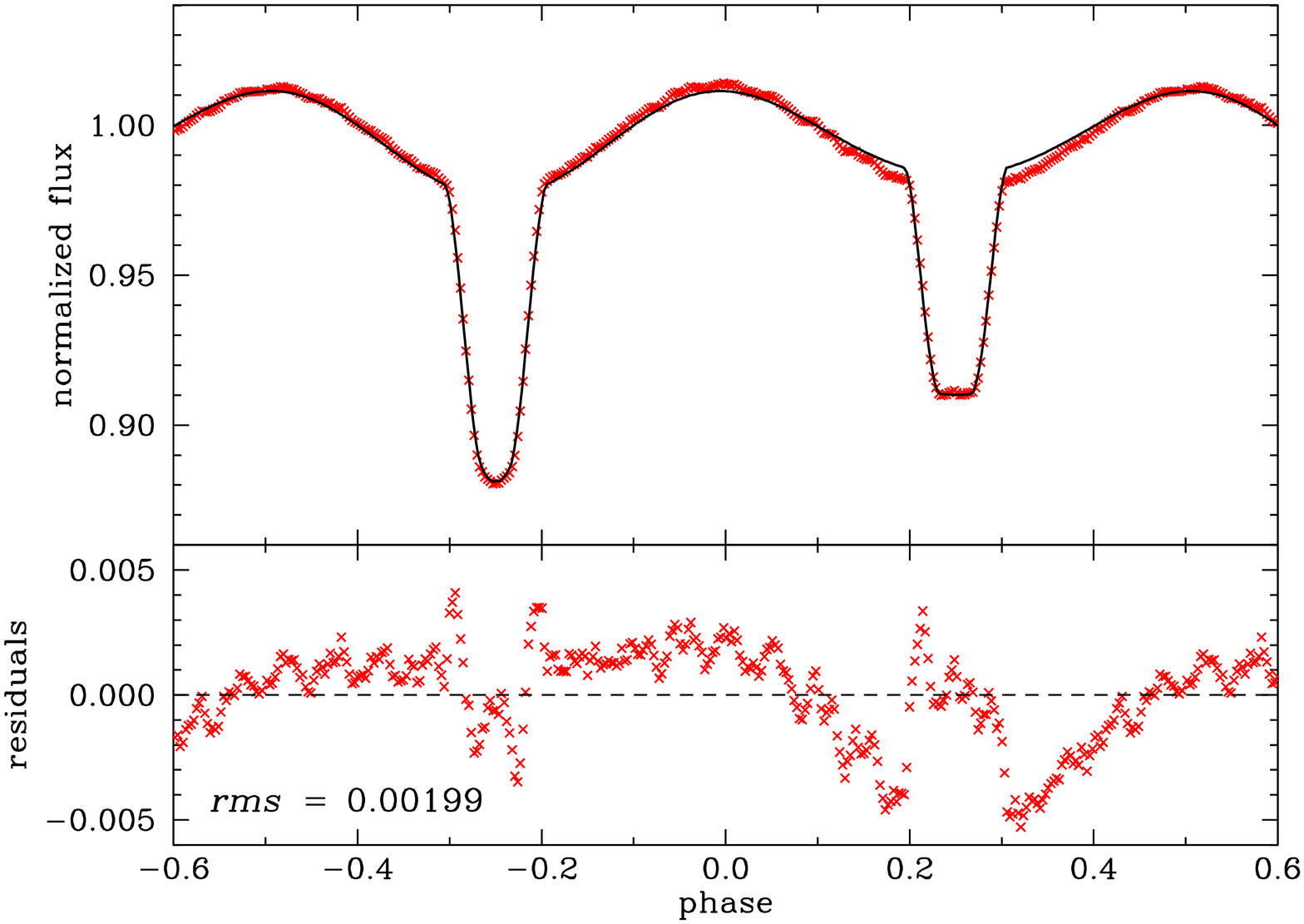}}
\end{minipage}
\begin{minipage}[b]{0.49\textwidth}
\flushright
(b)\hspace{0.5cm}\,\\[-4cm]
\centering
\resizebox{\hsize}{!}{\includegraphics[angle=-90]{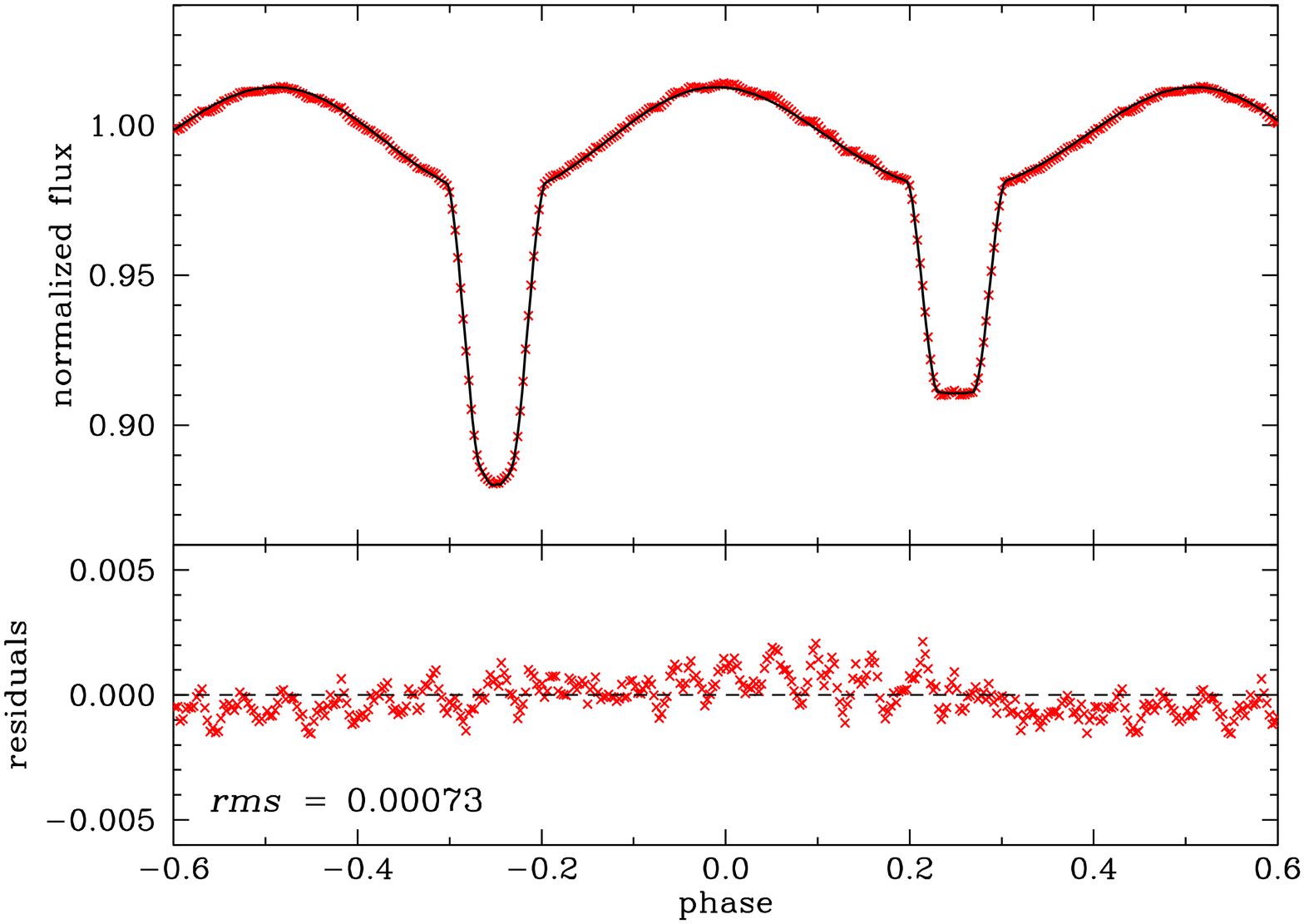}}
\end{minipage} \\
\begin{minipage}[b]{0.49\textwidth}
\flushright
(c)\hspace{0.5cm}\,\\[-4cm]
\centering
\resizebox{\hsize}{!}{\includegraphics[angle=-90]{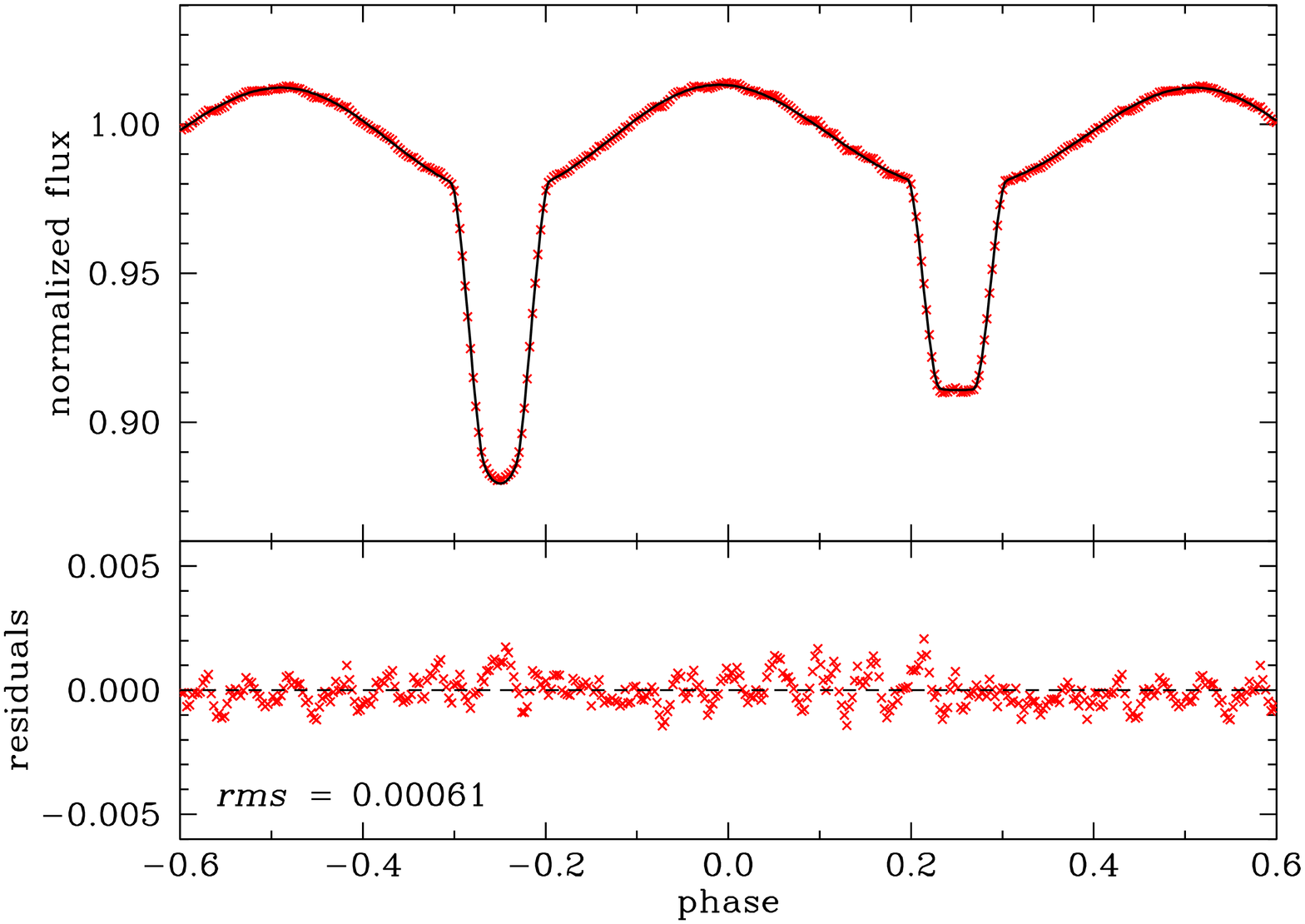}}
\end{minipage}
\begin{minipage}[b]{0.49\textwidth}
\flushright
(d)\hspace{0.5cm}\,\\[-4cm]
\centering
\resizebox{\hsize}{!}{\includegraphics[angle=-90]{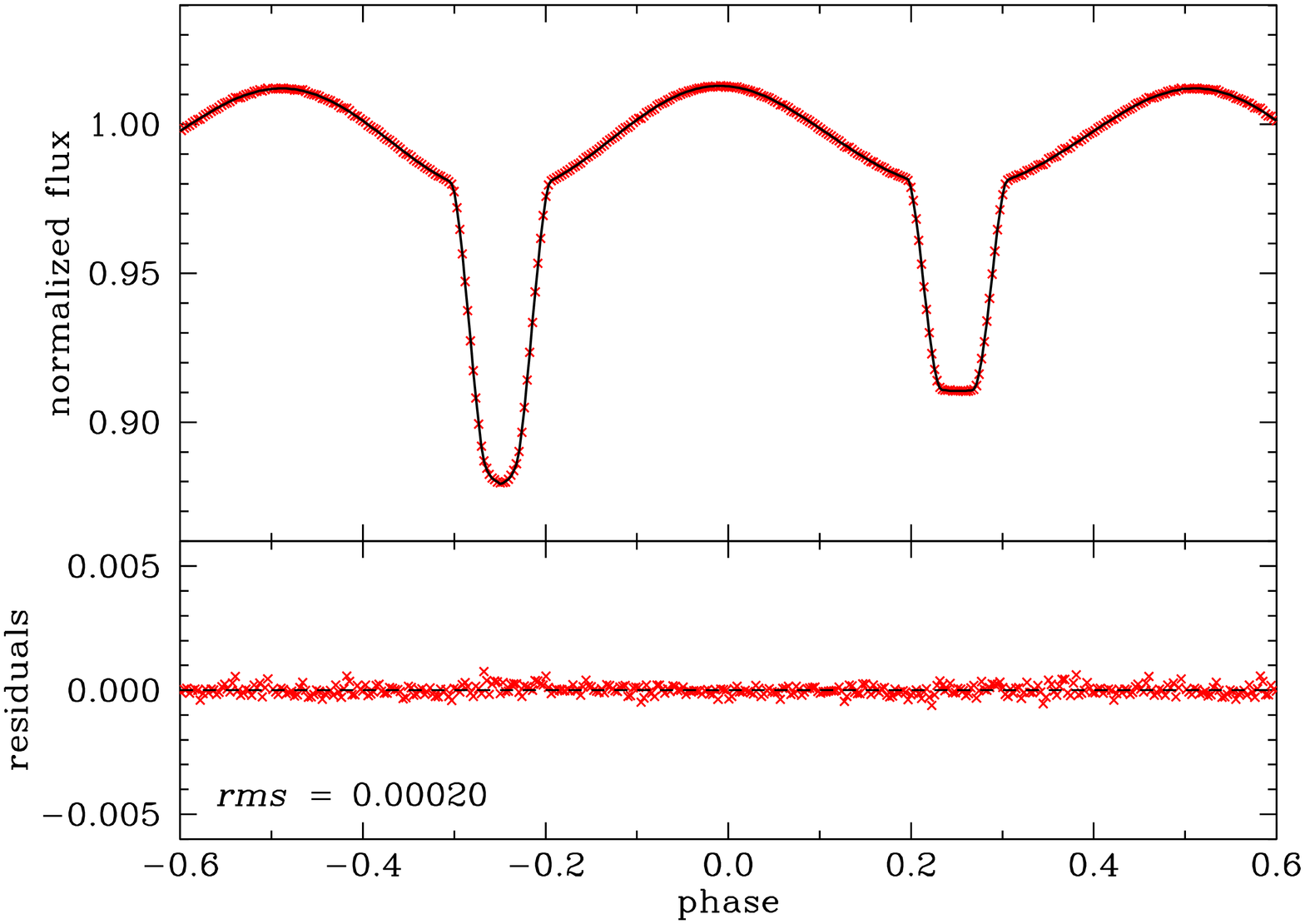}}
\end{minipage}
\caption{Phase-folded light curve ({\it the top of each panel}) and
	 residuals of the best-fit ({\it the bottom of each panel}) for
	 models (a) with gravity darkening and surface albedo coefficients
	 fixed at $\beta$ = 0.32 and $A$ = 0.60, (b) with $A_1$, $A_2$, and
	 $\beta_1$ set as free parameters but without accounting for the
	 Doppler beaming, (c) with $A_1$, $A_2$, $\beta_1$, and $B$ set as
	 free parameters but still keeping the oscillations (see
	 Sect.~\ref{pre_sol}), and (d) removing the oscillations.}
\label{lc_obs_ph}
\end{figure*}

For the spectral analysis of the primary component we assumed a flux ratio
of 0.07 that was estimated from the light curve solution (the ratio between
the bolometric luminosities listed in Table~\ref{par_tab}). We used the
ATLAS9 plane-parallel and LTE model atmospheres \citep{Kurucz1993}, and
calculated synthetic spectra using SPECTRUM, a stellar spectral synthesis
code \citep{GrayCorbally1994}. We convolved the synthetic spectra with a
Gaussian profile having a FWHM = 0.11~\AA, to account for the spectrograph
resolution, and adopted a surface gravity of \logg\ = 3.53 $\pm$ 0.01~dex,
as derived from the modeling of the eclipsing binary.

We derived the effective temperature by fitting the wings of the H$\alpha$
and H$\beta$ Balmer lines, and the iron abundance ([Fe/H]) and
micro-turbulence velocity (\vmicro) by applying the method described in
\citet{BlackwellShallis1979} on isolated \ion{Fe}{i} and \ion{Fe}{ii} lines.
The results are \teffp\ = 6750 $\pm$ 150~K, [Fe/H] = 0.0 $\pm$ 0.1~dex,
and \vmicrop\ = 2.5 $\pm$ 0.8~\kms. We also measured the projected
rotational velocity \vsini\ of the stars $i)$ by fitting the CCF profile of
individual FEROS spectra using the rotation profile described in
\citet{Gray1976}, which was convolved with the instrumental profile, and
$ii)$ by fitting the profile of several clean and unblended metal lines. We
derived \vsinip\ = 47.8 $\pm$ 0.5~\kms\ and \vsinis\ = 19 $\pm$ 3~\kms\ from
the CCF peaks for both primary and secondary stars, and \vsinip\ = 46 $\pm$
2~\kms\ from the metal lines of the decomposed spectrum of the primary. As a
matter of fact, the value of \vsini\ that we derive here is a mixed of
velocity fields including stellar rotation, macro turbulence, and
pulsations. However, given the large rotation rate of these stars,
macro-turbulence and pulsation broadening are second order effects.

We also derived the abundance of chemical elements other than iron using the
decomposed spectrum of the primary star. However, the continuum undulations
produced by the disentangling procedure, though we tried to correct them,
may still affect the abundance determination, specially for heavy elements,
which have a small number of spectral lines available. Nevertheless, for
elements lighter than Fe, the abundances mostly follow the solar values
\citep{Asplundetal2005}.

\begin{figure*}
\centering
\resizebox{\hsize}{!}{\includegraphics[angle=-90]{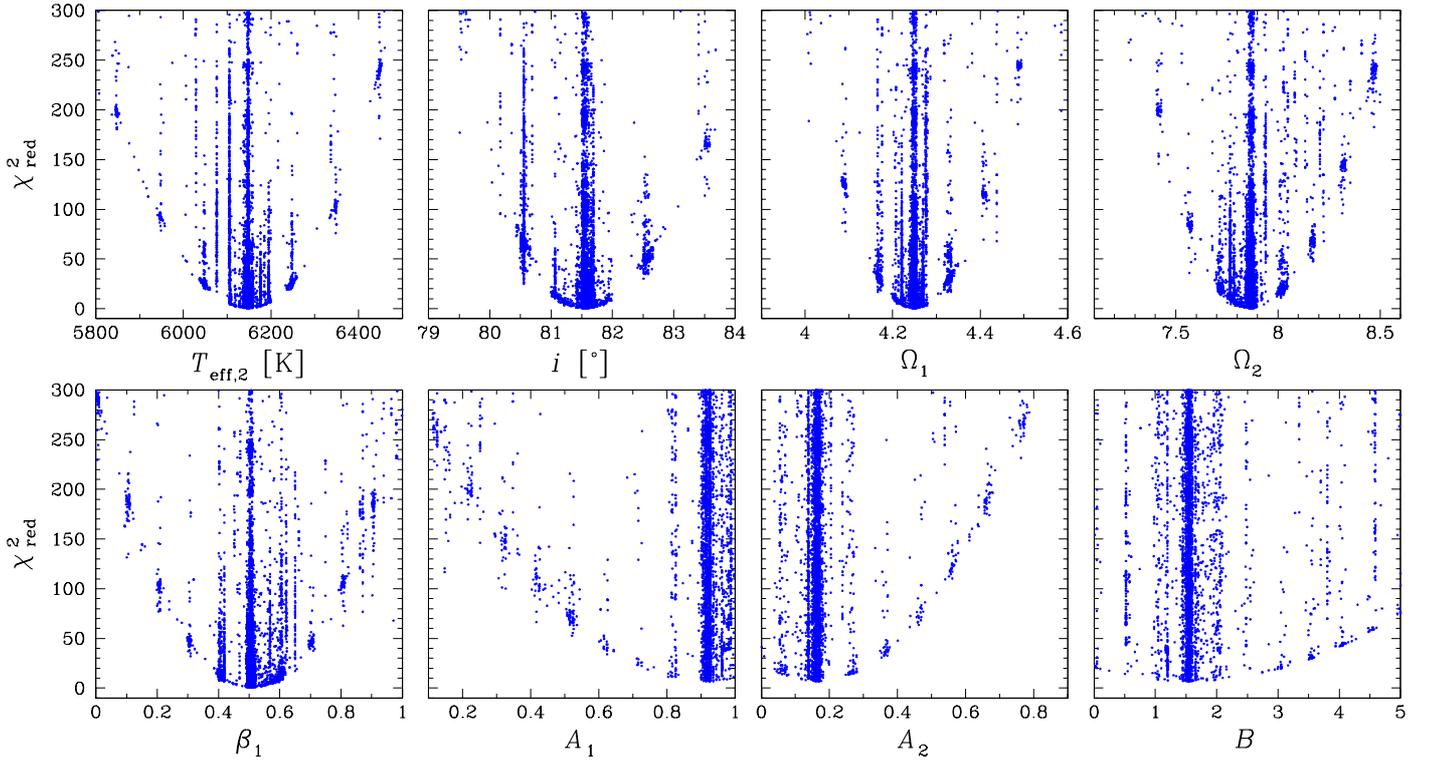}}
\caption{Reduced chi-square between model and data as a function of the
         eight free parameters of the light curve model computed using the
	 PIKAIA genetic algorithm during the search for the best solution.
	 From left to right, and from top to bottom: secondary effective
	 temperature, orbital inclination angle, surface potentials, primary
	 gravity darkening, surface albedos, and beaming factor.}
\label{pikaia}
\end{figure*}

%
%
\section{Light and radial-velocity curve analysis}
\label{analysis}

The binary model for both light and radial-velocity curve of
CoRoT\,105906206 were computed with PHOEBE, a tool for eclipsing binary
modeling based on the Wilson-Devinney code \citep{WilsonDevinney1971}. The
search for the best model was performed with the PIKAIA subroutine
\citep{Charbonneau2002}, a genetic-algorithm based approach. PIKAIA
maximizes a user-supplied function  ($= 1/\chi^2$) by minimizing the
chi-square $\chi^2$ between model and data. Starting with a so-called
``population", formed by a given number of ``individuals" which, in turn,
are formed by sets of parameters randomly selected within a given range of
the parameter space. These sets of parameters are encoded as strings to form
a ``chromosome-like" structure, and represent the first trial solutions. The
initial solutions (or ``parents") then evolve through subsequent
``generations" of possible solutions (the ``offsprings"), following rules
similar to typical processes of the biological evolution (breeding,
crossover, mutation). In equivalence to the natural selection, only
solutions that provide small values of $\chi^2$ are passed to the next
generations, and the final solution is achieved after some condition is
satisfied.

Our supplied function evaluates the $\chi^2$ between the observed light
curve and the models computed for a set of free parameters that characterize
the orbital and physical properties of a binary system. Following the same
procedure described in \citet{Maceronietal2014}, we implemented this
function in a FORTRAN-based routine, FITBINARY, which makes use of
Wilson-Devinney routines to compute the binary models. In our FITBINARY code
we adopt the PIKAIA 1.2 new version, an improvement of PIKAIA 1.0 that
includes additional genetic operators and algorithm strategies. In
particular, the new version allows several mutation modes such as a fixed or
an adjustable mutation rate based either on fitness (the value of 
$1/\chi^2$ comparing best and median individuals) or on distance (the metric
distance between best and median population clustering). We normally adopted
an initial population with 100 individuals evolving through 200 generations,
a mutation mode of one-point+creep and adjustable rate based on fitness, and
a full generation replacement with elitism. The other control parameters
were normally kept fixed at their default values (see the related
documentation in the PIKAIA
Homepage\footnote{http://www.hao.ucar.edu/modeling/pikaia/pikaia.php}).

In the prewhitening process, described in detail in Sect.~\ref{fin_sol},
$(i)$ we first remove an initial binary model from the processed light curve
(the one resulting from the cleaning steps of Sect.~\ref{phot}), then
$(ii)$ we identify the significant frequencies in the amplitude spectrum of
the time series with oscillations only,
$(iii)$ we subtract the frequencies from the processed light curve, and
finally
$(iv)$ we search for an improved solution for the binary model applied to
the light curve with the eclipses only.
These steps are repeated until there is no improvement on the last solution.
However, by using this procedure to treat the whole data set
(of about 144~days) we did not find any satisfying solution and the
remaining residuals still had clearly visible oscillations. This does not
happen if we apply the same procedure to a segment of light curve, since the
residuals in this case are at least 2.5 times smaller. Hence, we divided the
light curve into 8 segments of about 20~days each, and apply the procedure
of prewhitening to each one separately. In the following sections we present
the results obtained using the first segment of the light curve, and at the
end of the paper we comment on the comparison with what we obtain using the
other segments.

\begin{figure*}
\centering
\resizebox{\hsize}{!}{\includegraphics[angle=-90]{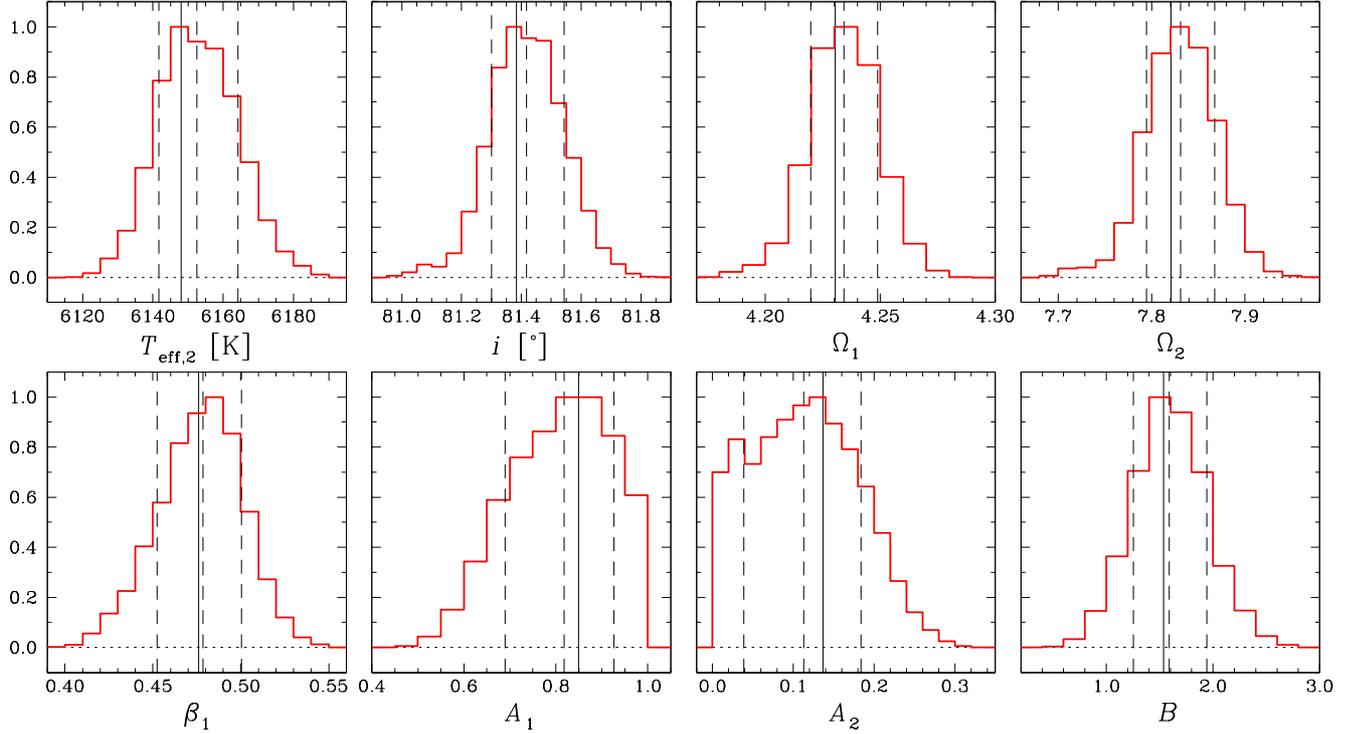}}
\caption{Distributions of the parameters resulted from our Markov chain
         Monte Carlo (MCMC) simulations. The histograms are normalized to
	 unit and contain 40,000 accepted solutions each. The median of each
	 distribution and the limits enclosing 68\% of the results (with
	 equal probability in both sides) are indicated by the dashed lines,
	 and represent the values listed in Table~\ref{par_tab}. The solid
	 lines indicate the solution that minimizes $\chi^2$.}
\label{mcmc_err}
\end{figure*}

\subsection{Preliminary solutions}
\label{pre_sol}

First, using PHOEBE and the radial-velocity data only, we searched for a
good solution to the orbital model by changing the orbital period $P$, the
semi-major axis $a$, the secondary-to-primary mass ratio $q$, and the
barycentric velocity \vga. Then, keeping fixed the derived values (except
$P$), we improved the orbital period and, with PIKAIA, we searched for a
solution that best fits the light curve data by changing the secondary
effective temperature \teffs, the orbital inclination angle $i$, and the
surface potentials $\Omega_1$ and $\Omega_2$. The eccentricity $e$ and the
longitude of periastron $\omega_0$ were kept fixed at zero, and the primary
effective temperature \teffp\ was set to 6800~K (according to our estimate
performed in Sect.~\ref{spec}). The ratio between orbital and rotation
periods was fixed to $f = 1.0$ for both companions. Later on, with our
estimate of the projected rotation velocity from the CCF peaks (see
Sect.~\ref{disent}), we derived \Protp\ = 4.44 $\pm$ 0.07~days and
\Prots\ = $3.63^{+0.53}_{-0.74}$~days (assuming coplanarity between
equatorial and orbital planes, as discussed in Sect.~\ref{remarks}), which
yielded $f_1$ = 0.83 $\pm$ 0.01 and $f_2$ = $1.02^{+0.15}_{-0.21}$. However,
these new values did not significantly affect the results already found for
the parameters.

Concerning the limb darkening, we adopted a square-root law with two
coefficients $x_{\rm LD}$ and $y_{\rm LD}$ for each component, which also
depend on the passband intensities. For a given set of photospheric
parameters, PHOEBE interpolates the limb darkening coefficient tables
computed for CoRoT passbands \citep[see][]{Maceronietal2009}. For both
stars, the bolometric luminosity $L$, mass $M$, radius $R$, and surface
gravity \logg\ were computed rather than adjusted, where $R$ represents the
radius of a sphere with the same volume as the modeled star.

This first preliminary solution model for a detached binary, which is shown
in Fig.~\ref{lc_obs_ph}a, is based on the cleaned light curve described in
Sect.~\ref{phot}, including the pulsations and other kind of periodic
patterns. The curve was binned using an average of 1000 points in both time
and flux in the phase-folded space. The gravity darkening and surface albedo
coefficients were kept fixed at $\beta$ = 0.32 and $A$ = 0.60 for purely
convective envelopes \citep{Lucy1967}, adopting
$T_{\rm eff}^4 \propto\ g^{\,\beta}$.

\citet{Claret1999} computed the gravity darkening as a function of mass and
degree of evolution of the star, which resulted in a smooth transition
between convective and radiative energy transport mechanisms. Therefore, in
a second step, the gravity darkening (except $\beta_2$) and surface albedos
were set as free parameters as a tentative to improve the fit (see
Fig.~\ref{lc_obs_ph}b). Since the primary luminosity dominates the observed
flux, the models are independent of the secondary gravity darkening, and
therefore $\beta_2$ was kept fixed (see Table~\ref{par_tab}).

As can be seen in Fig.~\ref{lc_obs_ph}b, more clearly in the residuals plot,
there is a modulation pattern, commonly referred to as the O'Connell effect
\citep{OConnell1951}, in which the flux maximum after the deeper eclipse is
larger than the maximum after the shallower eclipse
\citep[see also][]{Milone1968,DavidgeMilone1984}. Some attempted
explanations for this asymmetric modulation are the presence of spots on the
surface of a chromosphericaly active star, circumstellar clouds of gas and
dust, or a hot spot created by mass transfer. Given the physical
configuration of this system, however, both stars are nearly spherical (see
Sect.~\ref{phy_prop}) and thus the scenario of mass transfer can be safely
excluded from the list of possible causes of the O'Connell effect. Another
possible explanation instead is the so-called Doppler beaming, in which the
orbital movement of the star modifies its apparent luminosity, beaming the
emission of photons in the direction of the observer. The relation
\citep[adapted from][]{LoebGaudi2003} between emitted ($F_{0,\lambda}$) and
observed ($F_\lambda$) flux is:
\begin{equation}
\label{beam_eq}
F_\lambda = F_{0,_\lambda}\left(1 - B\,\frac{\upsilon_{\rm rad}}{c}\right)
\end{equation}
where $B$ is the beaming factor \citep[see also][]{Zuckeretal2007,
Bloemenetal2011}. $B$ depends on the wavelength of the observations and on
the physical characteristics of the star. In a binary system, since the
stars move in opposite directions with respect to the observer, the net
beaming effect represents the difference in flux between the components.
Therefore, there is no modulation caused by the Doppler beaming if the
binary components have the same mass and the same spectral type. In order to
account for and remove this modulation pattern from the observed light
curve, we searched for a binary model in which the emitted flux is
transformed according to Eq.~\ref{beam_eq} and the beaming factor is set as
a free parameter (see Fig.~\ref{lc_obs_ph}c and Table~\ref{par_tab}).

\subsection{Final solution, uniqueness, and uncertainties}
\label{fin_sol}


\begin{figure*}
\centering
\begin{minipage}[b]{0.33\textwidth}
\centering
\resizebox{\hsize}{!}{\includegraphics{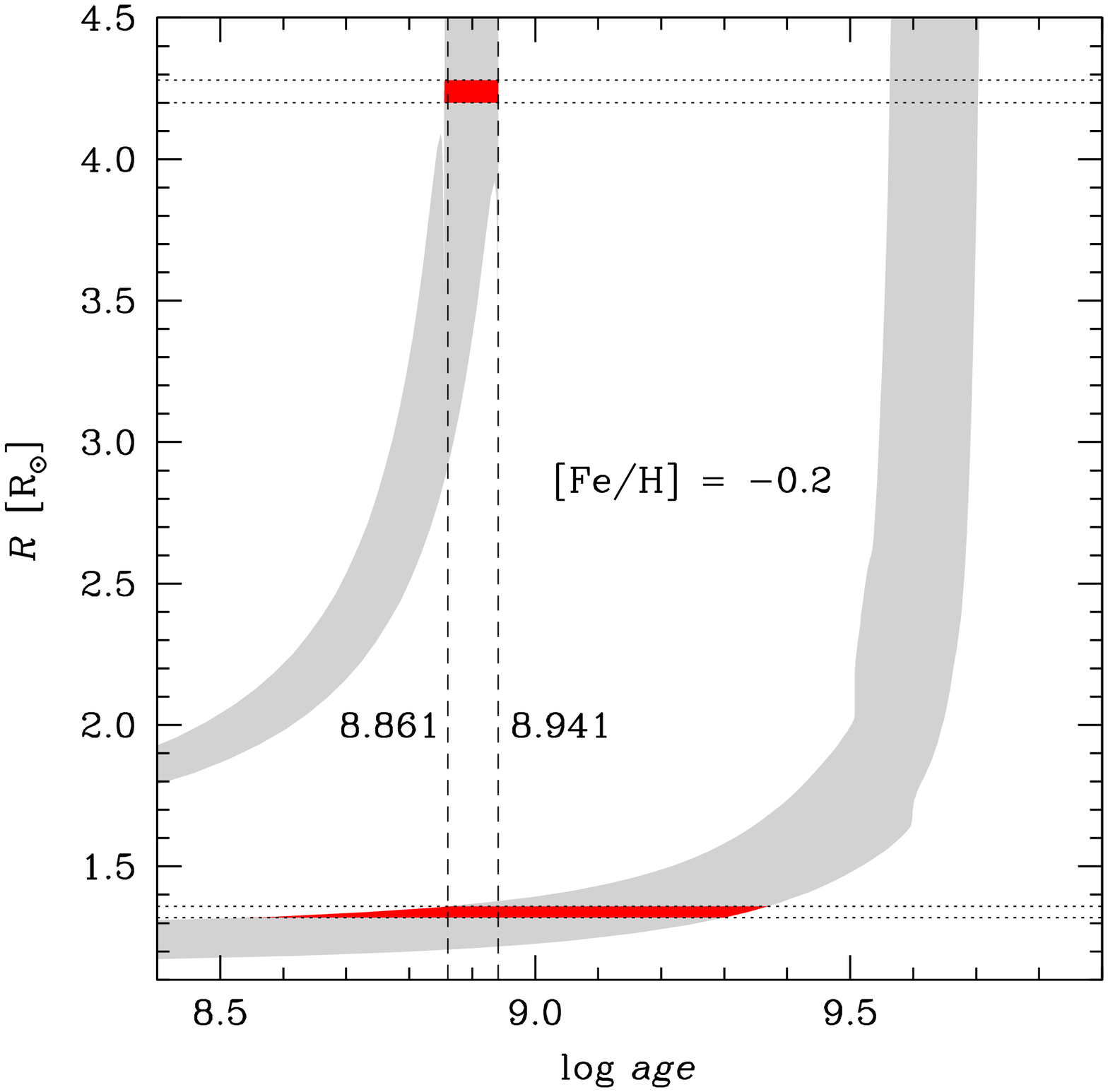}}
\end{minipage}
\begin{minipage}[b]{0.33\textwidth}
\centering
\resizebox{\hsize}{!}{\includegraphics{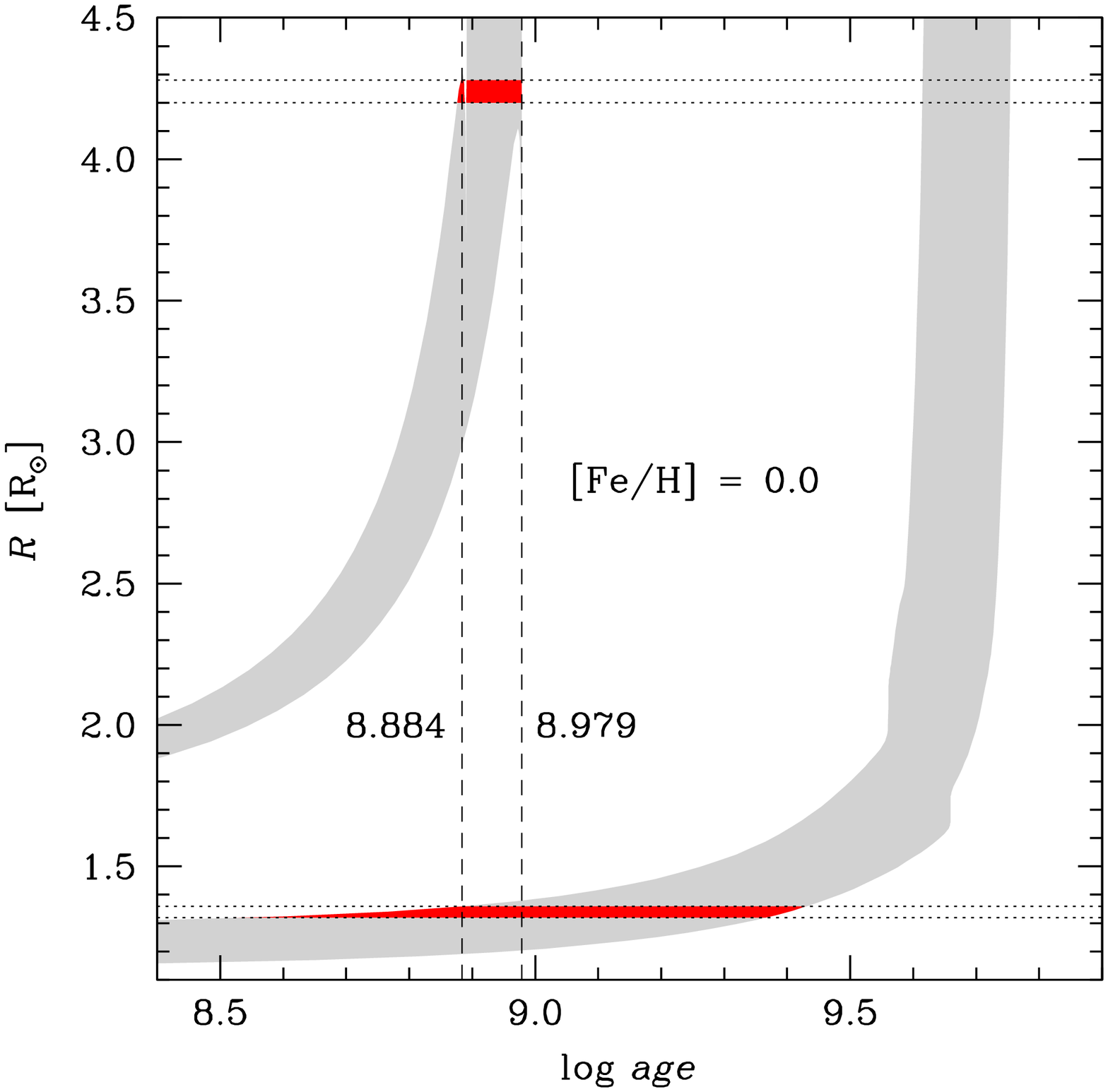}}
\end{minipage}
\begin{minipage}[b]{0.33\textwidth}
\centering
\resizebox{\hsize}{!}{\includegraphics{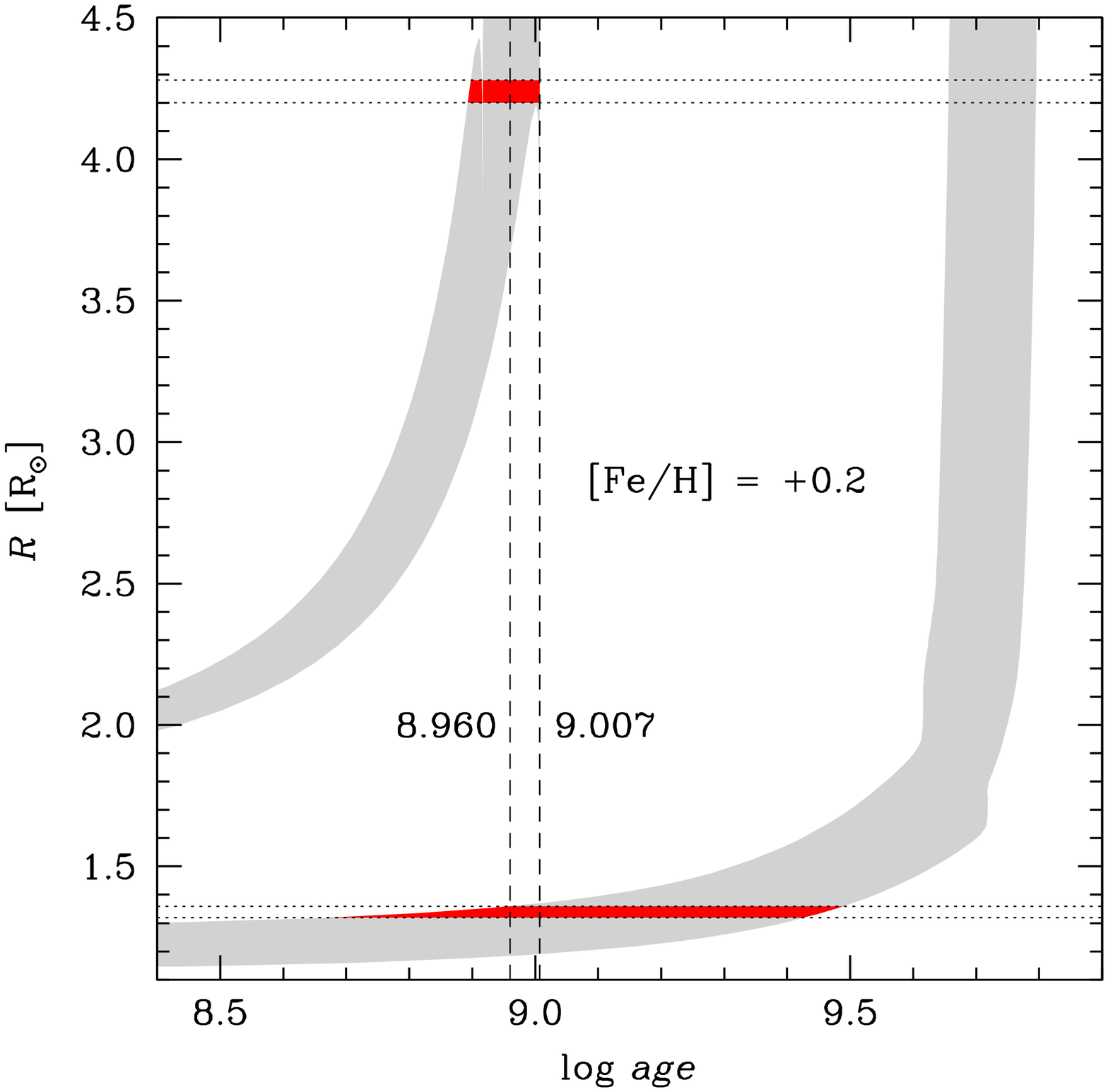}}
\end{minipage}
\caption{Evolution of the radius of the system components in the mass range
         $M_{1,2} \pm 2\sigma(M_{1,2})$ (shaded regions, from
	 Table~\ref{par_tab}) for three values of chemical composition. The
	 darker regions represent the intersection with $R_{1,2} \pm 
	 2\sigma(R_{1,2})$ (horizontal lines, also from
	 Table~\ref{par_tab}). The vertical lines indicate the limits in age
	 according to the constraint of coevality of the two stars.}
\label{tracks}
\end{figure*}

The plots in Figs.~\ref{lc_obs_ph}a, \ref{lc_obs_ph}b and \ref{lc_obs_ph}c
are based on the light curve with the oscillations still present. These
include those originating from stellar pulsations, and other sources such as
the orbital period of the CoRoT satellite (103~min.) and its harmonics. We
use an iterative procedure for the prewhitening. First we find the best fit
to the light curve setting the secondary effective temperature, orbital
inclination angle, surface potentials, the albedo coefficients of both
components, the primary gravity darkening, and the beaming factor as free
parameters (Fig.~\ref{lc_obs_ph}c). After that we subtract this model from
the observed data so that only oscillations remain.

We then use Period04 \citep{LenzBreger2005}, a program developed for the
analysis of time series, to compute the Fourier spectrum and to search for
the frequency, amplitude, and phase of each sinusoidal component of the
signal. To establish the number of components that give a significant
contribution to the signal, we apply the commonly-used criterion of
S/N $>$ 4, i.e., only frequencies having an amplitude 4 times the local
noise are considered. It is worth noting that, in order to avoid spurious
frequencies in the amplitude spectrum, small gaps in the time series were
filled in with artificial points that follow a polynomial function fitted to
adjacent regions of each gap. A noise consistent with the dispersion of the
adjacent regions was applied to each point. More details on the pulsation
frequencies are described in Sect.~\ref{puls_prop}.

Finally, the significant frequencies (according to the criterion mentioned
above) are subtracted from the cleaned light curve leaving only the
eclipses and the modulation of the binary. Again, using PIKAIA and PHOEBE,
we improve the preliminary values of \teffs, $i$, $\Omega_1$, $\Omega_2$,
$A_1$, $A_2$, $\beta_1$, and $B$. The procedure is repeated until the
$\chi^2$ value between model and data does not decrease significantly.

We also tried to account for the individual flux contribution of each star
to the observed pulsations during and out of the eclipses. However, since
the primary star is the pulsating component (we will get to this conclusion
in Sect.~\ref{puls_prop}) and dominates the observed flux, the influence of
the secondary component on the observed pulsations is negligible, being
smaller than the noise. We were thus unable to see any difference in the
pulsations amplitude in- and out-of-eclipses, and the pulsation frequencies
found were equally subtracted from the entire light curve.

The minimization algorithms employed by PHOEBE are the commonly used
Levenberg-Marquandt and Nelder \& Mead's downhill simplex, both nonlinear
optimization methods affected by the problem of remaining stuck at local
minima. In the PIKAIA genetic algorithm, mutation and crossover allow the
population to move away from local solutions. Figure~\ref{pikaia} plots the
reduced chi-square $\chi^2_{\rm red} = \chi^2/\nu$ (where $\nu = N-n-1$ is
the number of degrees of freedom for $N$ observation points and $n$ fitted
parameters) as a function of the free parameters of the light curve model,
for a wide range of the parameter space. Note that PIKAIA tries several
local solutions (pattern of points vertically assembled), and progressively
moves towards the global minimum. In any case, we preferred to carry out
more than one iteration, and also around a shorter range of the parameter
space, to better constrain the solutions to provide the minimum $\chi^2$.

\begin{table}
\centering
\caption[]{Parameters of CoRoT\,105906206 from PHOEBE fits of light and
           radial-velocity curves.}
\label{par_tab}
\begin{tabular}{l r@{}l r@{}l}
\hline\hline\noalign{\smallskip}
 & \multicolumn{4}{c}{system} \\
 & \multicolumn{2}{c}{primary} & \multicolumn{2}{c}{secondary} \\
\noalign{\smallskip}\hline\noalign{\smallskip}
$P$ [days]    & \multicolumn{4}{c}{3.69457080 $\pm$ 0.00000013} \\
$a$ [\Rsun]   & \multicolumn{4}{c}{15.32 $\pm$ 0.08\,\,\,} \\
$q = M_2/M_1$ & \multicolumn{4}{c}{\,0.574 $\pm$ 0.008\,} \\   
\vga\  [\kms] & \multicolumn{4}{c}{$-$30.30 $\pm$ 0.34\,\,\,\,\,\,\,} \\
$e$           & \multicolumn{4}{c}{0 (fixed)} \\   
$\omega_0$    & \multicolumn{4}{c}{0 (fixed)} \\   
$i$ [\degr]   & \multicolumn{4}{c}{81.42 $\pm$ 0.13\,\,\,} \\   
$B$           & \multicolumn{4}{c}{1.59 $\pm$ 0.35} \\
\teff\  [K]   & 6750 &\ $\pm$ 150  & 6152 &\ $\pm$ 162 \\
$\Omega$      & 4.23 &\ $\pm$ 0.01 & 7.83 &\ $\pm$ 0.04 \\
$\beta$       & 0.48 &\ $\pm$ 0.03 & \multicolumn{2}{c}{0.32 (fixed)} \\
$A$           & 0.82 &\ $\pm$ 0.13 & 0.11 &\ $\pm$ 0.07 \\
\xld          & \multicolumn{2}{c}{0.249747} & \multicolumn{2}{c}{0.286623} \\
\yld          & \multicolumn{2}{c}{0.474672} & \multicolumn{2}{c}{0.441072} \\
\logL         & 1.53 &\ $\pm$ 0.04 & 0.36 &\ $\pm$ 0.05 \\
$M$ [\Msun]   & 2.25 &\ $\pm$ 0.04 & 1.29 &\ $\pm$ 0.03 \\
$R$ [\Rsun]   & 4.24 &\ $\pm$ 0.02 & 1.34 &\ $\pm$ 0.01 \\
\logg         & 3.53 &\ $\pm$ 0.01 & 4.30 &\ $\pm$ 0.02 \\
\hline
\end{tabular}
\end{table}

\begin{figure*}
\centering
\resizebox{\hsize}{!}{\includegraphics[angle=-90]{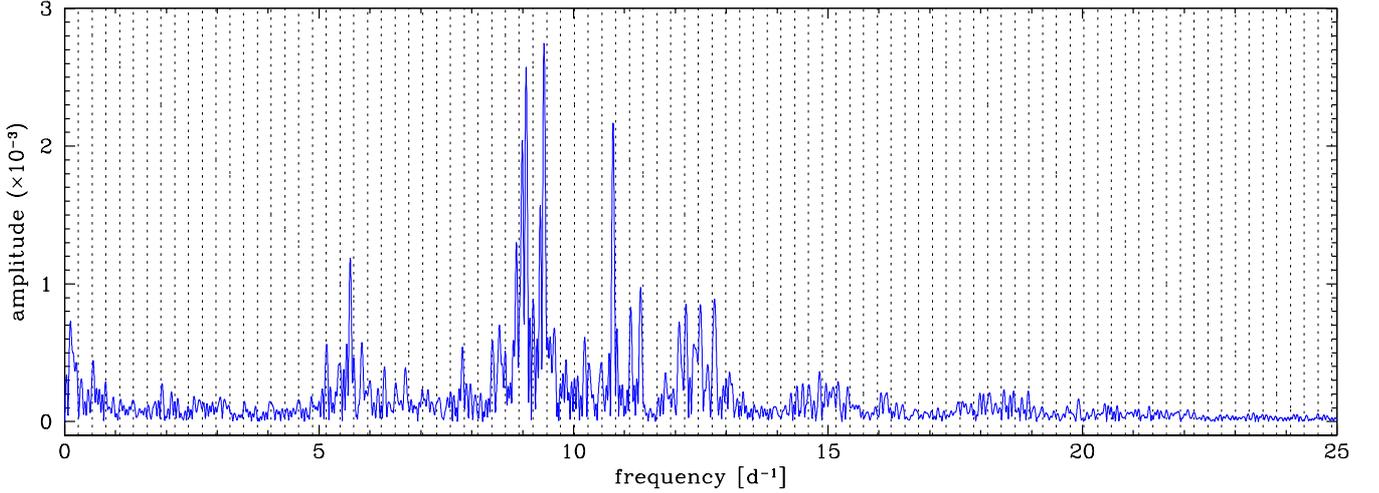}}
\caption{Amplitude spectrum of CoRoT\,105906206 after subtracting the
         final binary model. The vertical dashed lines represent multiples
	 of the orbital frequency (\Forb\ = 0.270667~d$^{-1}$).}
\label{freq_spec}
\end{figure*}

To estimate the uncertainties in the parameters we performed Markov chain
Monte Carlo (MCMC) simulations using the Metropolis algorithm
\citep[see e.g.][]{Gelmanetal2003,Ford2005}. This method is based on a
bayesian inference in the sense that the posterior (or desired) probability
distribution $p$($\bm{x} {\mid} \bm{d}$) of a set of parameters $\bm{x}$
given the observed data $\bm{d}$ is proportional to the product of a prior
knowledge of the probability distribution of the parameters $p$($\bm{x}$)
and a factor $\exp{[-\chi^2(\bm{x})/2]}$, which depends on the $\chi^2$
goodness of the fit adopting the model $\bm{x}$. For the prior probability
function we adopted a Gaussian distribution, centered around an initial set
of $\bm{x}$ and having initial arbitrary variances. The idea is to construct
a parameter space that approximates the desired probability distribution. To
do so, we generate a chain (or sequence) formed by sets of random parameters
that are sampled according to the prior probability distributions. If a set
of parameters $\bm{x'}$ yields a $\chi^2$ smaller than the previous one,
then the new solution is accepted. Otherwise, this new solution is accepted
with probability $\exp{[-[\chi^2(\bm{x'}) - \chi^2(\bm{x})/2]]}$.

The acceptance rate can be adjusted by multiplying the initial variances by
a scaling factor, in order to optimize the number of accepted solutions. We
redefine this scaling factor every 100 steps of the simulations to achieve a
ratio of about 0.25 between the accepted and the total number of generated
solutions, which is an optimal value for a multi-dimension space
\citep[see][]{Gelmanetal2003}. During a first phase of tests, which is
called the burn-in phase, the variances of the prior distributions are
constantly updated, being computed from the posterior distributions
themselves. The uncertainty on the data is also adjusted to ensure that the
value of minimum $\chi^2$ is equal to the number of degrees of freedom. The
burn-in solution is then discarded and a new simulation begins. After
reaching a certain number of accepted solutions, the posterior probability
distributions become stable and the chain converges towards the desired
distribution.

Figure~\ref{mcmc_err} plots the distributions of the eight free parameters
that we used to model the light curve of CoRoT\,105906206. We discarded the
first 15\% of the accepted solutions and the histograms contains 40,000
points each. The parameters of the best-fit solution and their uncertainties
are listed in Table~\ref{par_tab}. They represent the median values and the
68\% confidence limits shown in the figure, and are in very good agreement
with the best solution found with PIKAIA. It is worth noting that, since
\teffs\ is correlated with \teffp\  (because of the degeneracy between the
effective temperatures and the passband luminosities), the uncertainty
estimated for the secondary temperature also accounts for the one estimated
for the primary star. The final light curve model is shown in
Fig.~\ref{lc_obs_ph}d.

%
%
\section{Physical properties of CoRoT\,105906206}
\label{phy_prop}

Besides the parameters that come directly from the fits of the light and
radial-velocity curves, the masses ($M_1$ and $M_2$), radii ($R_1$ and
$R_2$), and surface gravities (\loggp\ and \loggp) of the two stars are also
calculated. Using the values of radius and effective temperature, we can
calculate the luminosities ($L_2$ and $L_1$). All these parameters, together
with the limb darkening coefficients (\xldp, \xlds, \yldp, \ylds), are
listed in Table~\ref{par_tab}. The resulting model is for a detached binary,
whose components are both nearly spherical. For the primary, the radius in
the direction toward the secondary is about 4\% larger than the radius in
the direction of the stellar pole. For the secondary this value is only
about 0.3\%. This shows that even the primary star is far from reaching the
Roche lobe.

In order to estimate the age of the system, we used the Yonsei-Yale (Y$^2$)
evolutionary tracks \citep{Yietal2003}, interpolated in mass and chemical
composition, drawn on the radius vs. age diagram. This kind of diagram
allows us to see the evolution of the stellar radius with time, and is
preferred over using the placing of the stars in the H-R diagram because in
eclipsing binary systems we have a precise determination of mass and radius
of the components. We checked the effective temperatures of each star that
come from the evolutionary models and they are consistent with those derived
from our spectral and light curve analyses.

Figure~\ref{tracks} shows the radius vs. age diagrams for three values of
metallicity in the range [Fe/H] $\pm$ 2~$\sigma$([Fe/H]), from
Sect.~\ref{disent}, plotted for masses and radii in the range
$M_{1,2} \pm 2\sigma(M_{1,2})$ and $R_{1,2} \pm 2\sigma(R_{1,2})$, both from
Table~\ref{par_tab}. Applying the constraint of coevality of the two stars,
we found the age of this binary system to be $859^{+79}_{-67}$~Myr. These
are 1$\sigma$ uncertainties though, for clarity in the figure, the diagrams
were constructed within two standard deviations in metallicity, mass, and
radius. The sharp vertical drops correspond to the second gravitational
contraction at the end of main-sequence stars. To compute the evolutionary
tracks, an overshoot parameter $\alpha_{\rm OV}$ = 0.2 ${\rm H}_{\rm P}$
(units of the pressure scale height) is adopted when a convective core
develops \citep[see][and references therein]{Yietal2003}.

%
%
\section{Pulsation properties}
\label{puls_prop}

Figure~\ref{freq_spec} shows the amplitude spectrum of CoRoT\,105906206,
computed using Period04 after subtracting the final binary model from the
light curve. We searched for significant peaks up to the Nyquist frequency,
but none was found beyond 25~d$^{-1}$. A total of 220 frequencies with
S/N $>$ 4 was identified, and Table~\ref{freq} lists the first 50, together
with the amplitudes and phases, and the respective formal errors computed
according to \citet{MontgomeryODonoghue1999}. We also searched for possible
frequency combinations, including also the orbital frequency
(\Forb\ = 0.270667~d$^{-1}$). The most evident are listed in the last column
of this table. We identified a few combinations involving the orbital
frequency and the most dominant peaks $F_1$, $F_2$, $F_3$, and $F_5$, which
seem to be genuine p-mode frequencies. A few overtones of the orbital period
are still present in Fig.~\ref{freq_spec}, and are also indicated in
Table~\ref{freq}. They are probably residuals of the prewhitening process,
but have small amplitudes.

The typical properties of $\delta$~Sct type variables in terms of effective
temperature, luminosity, mass, evolutionary stage, and pulsation frequency
range are all present in the primary component of CoRoT\,105906206, as one
can see in Table~\ref{par_tab} and in the list of dominant frequencies in
Table~\ref{freq}. As an rough estimate of the frequency range of the
pulsations, we applied the relation between the frequency of a given radial
mode $F_{\rm puls}$, the stellar mean density $\overline{\rho}$, and the
pulsation constant $Q$:
\begin{equation}
Q\,F_{\rm puls} = \sqrt{\frac{\overline{\rho}}{\rho_{\sun}}}
\end{equation}
From the fundamental to the fifth radial mode we have
$0.033 \lesssim Q \lesssim 0.013$ \citep{Stellingwerf1979}, which yields
$5.2 \lesssim F_{\rm puls} \lesssim 13.2~{\rm d}^{-1}$ if we use the values
of mass and radius of the primary star, a range that contains the most
significant frequencies seen in Fig.~\ref{freq_spec}. The same computation
for the secondary star yields frequencies in the range
$22 \lesssim F_{\rm puls} \lesssim 56~{\rm d}^{-1}$, which are not observed
in the amplitude spectrum. This confirms that the identified pulsation modes
belong the primary component.

From the theoretical point of view, a non-adiabatic analysis of an
equilibrium model with the physical properties of the primary component,
performed with the non-adiabatic oscillation code MAD
\citep{Dupretetal2005}, provides excited modes with frequencies in the
typical domain of $\delta$~Sct type stars, namely:
from 5.2 to 17.0~${\rm d}^{-1}$ in the fundamental mode ($\ell = 0$),
from 4.0 to 16.6~${\rm d}^{-1}$ in the $\ell = 1$ mode, and
from 4.2 to 16.7~${\rm d}^{-1}$ in the $\ell = 2$ mode,
which are in line with the observed range of pulsation frequencies.

%
%
\section{Final remarks and conclusions}
\label{remarks}

The study of CoRoT\,105906206 unveiled an eclipsing binary system in which
the primary component pulsates in the range of frequencies typical $\delta$
Sct type variables, a result that agrees with the derived values of
effective temperature, luminosity, mass, and stage of evolution. Moreover,
from the theoretical point of view, a non-adiabatic analysis of a model
matching the physical properties of the primary star gives excited
frequencies of the most relevant modes ($\ell$ = 0, 1, and 2) in the same
frequency range of the observed pulsations.

By allowing the surface albedos and gravity darkening to vary as free
parameters we have improved the solution for the binary model in comparison
with the model found by keeping them fixed. In particular, our estimate of
$\beta_1$ is in agreement with the calculations of \citet{Claret1999} for
the gravity darkening as a function of the effective temperature for a
2~\Msun\ star.

An interesting characteristic of this system is the presence of the
O'Connell effect, an asymmetric photometric modulation pattern that we
interpreted as due to the Doppler beaming of the emitted flux. We have
quantified this effect by means of the beaming factor $B$, whose value
depends on the passband of the photometric observations and on the physical
properties of the stars. Using equations from \citet{MazehFaigler2010}, our
Eq.~\ref{beam_eq}, and some of the parameters in Table~\ref{par_tab}, we
obtain $B$ = 2.00 $\pm$ 0.05, which is in fair agreement with the beaming
factor derived from the light curve model (see Table~\ref{par_tab}). The
$\alpha_{\rm beam}$ factor used in those equations was set to unity. This
factor accounts for the effect of the stellar light being shifted out or
into the observed passband. According to \citet{FaiglerMazeh2011}, the value
of $\alpha_{\rm beam}$ may range from 0.8 to 1.2 for F, G, and K stars
observed with the CoRoT passband. The same beaming factor listed in
Table~\ref{par_tab} is obtained setting $\alpha_{\rm beam}$ = 0.8.

\begin{table*}
\centering
\caption[]{First 50 pulsation frequencies, amplitudes, and phases derived
           for CoRoT\,105906206 after subtracting the final binary model.}
\label{freq}
\begin{tabular}{l r@{}l c c c c}
\hline\hline\noalign{\smallskip}
 & \multicolumn{2}{c}{frequency [d$^{-1}$]} & amplitude $\times$ 10$^3$ &
phase [$2\pi$] & remark \\
\noalign{\smallskip}\hline\noalign{\smallskip}
$F_{1}$  &  9.&4175 $\pm$ 0.0001 & 2.552 $\pm$ 0.012 & 0.559 $\pm$ 0.005 & \\
$F_{2}$  &  9.&0696 $\pm$ 0.0001 & 2.296 $\pm$ 0.012 & 0.491 $\pm$ 0.005 & \\
$F_{3}$  & 10.&7776 $\pm$ 0.0002 & 2.150 $\pm$ 0.012 & 0.995 $\pm$ 0.006 & \\
$F_{4}$  &  8.&9951 $\pm$ 0.0002 & 1.913 $\pm$ 0.012 & 0.384 $\pm$ 0.006 & 2$F_2$ $-$ $F_1$ + \Forb \\
$F_{5}$  &  5.&6119 $\pm$ 0.0003 & 1.160 $\pm$ 0.012 & 0.313 $\pm$ 0.010 & \\
$F_{6}$  &  9.&3471 $\pm$ 0.0003 & 1.179 $\pm$ 0.012 & 0.809 $\pm$ 0.010 & $F_2$ + \Forb \\
$F_{7}$  &  8.&8734 $\pm$ 0.0003 & 1.051 $\pm$ 0.012 & 0.019 $\pm$ 0.011 & $F_1$ $-$ 2\Forb \\
$F_{8}$  & 11.&3164 $\pm$ 0.0004 & 0.903 $\pm$ 0.012 & 0.221 $\pm$ 0.013 & $F_3$ + 2\Forb \\
$F_{9}$  & 12.&7668 $\pm$ 0.0004 & 0.976 $\pm$ 0.012 & 0.701 $\pm$ 0.012 & 2$F_3$ $-$ $F_2$ + \Forb \\
$F_{10}$ & 12.&2116 $\pm$ 0.0004 & 0.832 $\pm$ 0.012 & 0.405 $\pm$ 0.014 & 2$F_3$ $-$ $F_2$ $-$ \Forb \\
$F_{11}$ &  8.&9203 $\pm$ 0.0004 & 0.906 $\pm$ 0.012 & 0.800 $\pm$ 0.013 & \\
$F_{12}$ & 12.&4963 $\pm$ 0.0004 & 0.967 $\pm$ 0.012 & 0.684 $\pm$ 0.012 & 2$F_3$ $-$ $F_2$ \\
$F_{13}$ &  9.&6056 $\pm$ 0.0004 & 0.806 $\pm$ 0.012 & 0.229 $\pm$ 0.015 & $F_2$ + 2\Forb \\
$F_{14}$ & 11.&1236 $\pm$ 0.0005 & 0.723 $\pm$ 0.012 & 0.325 $\pm$ 0.016 & $F_1$ $-$ $F_2$ + $F_3$ \\
$F_{15}$ &  0.&1192 $\pm$ 0.0004 & 0.880 $\pm$ 0.012 & 0.686 $\pm$ 0.013 & \\
$F_{16}$ &  9.&4713 $\pm$ 0.0005 & 0.715 $\pm$ 0.012 & 0.427 $\pm$ 0.017 & \\
$F_{17}$ & 12.&4186 $\pm$ 0.0005 & 0.718 $\pm$ 0.012 & 0.810 $\pm$ 0.017 & 2$F_3$ $-$ $F_1$ + \Forb \\
$F_{18}$ & 12.&0733 $\pm$ 0.0005 & 0.721 $\pm$ 0.012 & 0.170 $\pm$ 0.016 & \\
$F_{19}$ &  8.&5391 $\pm$ 0.0005 & 0.647 $\pm$ 0.012 & 0.356 $\pm$ 0.018 & $F_2$ $-$ 2\Forb \\
$F_{20}$ &  9.&2153 $\pm$ 0.0005 & 0.688 $\pm$ 0.012 & 0.874 $\pm$ 0.017 & \\
$F_{21}$ &  5.&8385 $\pm$ 0.0006 & 0.581 $\pm$ 0.012 & 0.229 $\pm$ 0.020 & \\
$F_{22}$ &  5.&1446 $\pm$ 0.0006 & 0.548 $\pm$ 0.012 & 0.441 $\pm$ 0.022 & 19\Forb \\
$F_{23}$ & 10.&2152 $\pm$ 0.0006 & 0.530 $\pm$ 0.012 & 0.259 $\pm$ 0.022 & \\
$F_{24}$ &  0.&1467 $\pm$ 0.0004 & 0.772 $\pm$ 0.012 & 0.672 $\pm$ 0.015 & \\
$F_{25}$ &  7.&8192 $\pm$ 0.0007 & 0.476 $\pm$ 0.012 & 0.445 $\pm$ 0.025 & \\
$F_{26}$ &  8.&4036 $\pm$ 0.0006 & 0.606 $\pm$ 0.012 & 0.902 $\pm$ 0.020 & \\
$F_{27}$ & 10.&3200 $\pm$ 0.0007 & 0.473 $\pm$ 0.012 & 0.258 $\pm$ 0.025 & 2$F_1$ $-$ $F_2$ + 2\Forb \\
$F_{28}$ & 12.&3571 $\pm$ 0.0008 & 0.435 $\pm$ 0.012 & 0.378 $\pm$ 0.027 & \\
$F_{29}$ &  0.&5590 $\pm$ 0.0007 & 0.478 $\pm$ 0.012 & 0.116 $\pm$ 0.025 & 2\Forb \\
$F_{30}$ &  8.&6016 $\pm$ 0.0008 & 0.426 $\pm$ 0.012 & 0.487 $\pm$ 0.028 & $F_1$ $-$ 3\Forb \\
$F_{31}$ &  5.&4000 $\pm$ 0.0008 & 0.419 $\pm$ 0.012 & 0.989 $\pm$ 0.028 & \\
$F_{32}$ & 12.&1453 $\pm$ 0.0008 & 0.448 $\pm$ 0.012 & 0.452 $\pm$ 0.026 & 2$F_3$ $-$ $F_1$ \\
$F_{33}$ & 12.&6386 $\pm$ 0.0009 & 0.397 $\pm$ 0.012 & 0.383 $\pm$ 0.030 & \\
$F_{34}$ &  0.&2011 $\pm$ 0.0010 & 0.349 $\pm$ 0.012 & 0.676 $\pm$ 0.034 & $F_2$ $-$ $F_1$ + 2\Forb \\
$F_{35}$ &  9.&5168 $\pm$ 0.0006 & 0.609 $\pm$ 0.012 & 0.777 $\pm$ 0.019 & \\
$F_{36}$ &  6.&6930 $\pm$ 0.0009 & 0.369 $\pm$ 0.012 & 0.856 $\pm$ 0.032 & \\
$F_{37}$ &  5.&5582 $\pm$ 0.0010 & 0.351 $\pm$ 0.012 & 0.785 $\pm$ 0.034 & \\
$F_{38}$ & 14.&8369 $\pm$ 0.0010 & 0.341 $\pm$ 0.012 & 0.855 $\pm$ 0.035 & \\
$F_{39}$ &  6.&2852 $\pm$ 0.0011 & 0.321 $\pm$ 0.012 & 0.361 $\pm$ 0.037 & \\
$F_{40}$ & 12.&8082 $\pm$ 0.0009 & 0.399 $\pm$ 0.012 & 0.037 $\pm$ 0.030 & \\
$F_{41}$ & 11.&8103 $\pm$ 0.0009 & 0.379 $\pm$ 0.012 & 0.440 $\pm$ 0.031 & \\
$F_{42}$ &  5.&9771 $\pm$ 0.0010 & 0.347 $\pm$ 0.012 & 0.606 $\pm$ 0.034 & $F_3$ $-$ $F_5$ + 3\Forb \\
$F_{43}$ & 10.&5627 $\pm$ 0.0011 & 0.308 $\pm$ 0.012 & 0.771 $\pm$ 0.038 & \\
$F_{44}$ &  9.&3057 $\pm$ 0.0009 & 0.401 $\pm$ 0.012 & 0.201 $\pm$ 0.030 & \\
$F_{45}$ & 10.&5060 $\pm$ 0.0012 & 0.292 $\pm$ 0.012 & 0.673 $\pm$ 0.041 & \\
$F_{46}$ &  5.&6726 $\pm$ 0.0011 & 0.321 $\pm$ 0.012 & 0.447 $\pm$ 0.037 & \\
$F_{47}$ & 13.&0780 $\pm$ 0.0010 & 0.345 $\pm$ 0.012 & 0.836 $\pm$ 0.034 & \\
$F_{48}$ &  8.&6659 $\pm$ 0.0011 & 0.317 $\pm$ 0.012 & 0.352 $\pm$ 0.037 & \\
$F_{49}$ &  0.&2962 $\pm$ 0.0013 & 0.268 $\pm$ 0.012 & 0.925 $\pm$ 0.044 & \\
$F_{50}$ & 15.&2025 $\pm$ 0.0012 & 0.299 $\pm$ 0.012 & 0.655 $\pm$ 0.040 & \\
\hline
\end{tabular}
\tablefoot{The uncertainties are the formal values computed using equations
	   from \citet{MontgomeryODonoghue1999}. The remark column shows the
	   most relevant frequency combinations (\Forb\ =
	   0.270667~d$^{-1}$). The whole table containing the 220 pulsation
	   frequencies is available in electronic form at the CDS.}
\end{table*}

The fact the primary star rotates with a sub-synchronous velocity may
generate some doubt on our determination of the rotation period, and
motivates us to provide some tentative explanations. A spin-orbit
misalignment, for example, would lead to an overestimation of \Protp.
However, using Eq. 22 of \citet{Hut1981} applied to the parameters derived
for this system, and assuming that the primary component rotates as a rigid
body, we can estimate the ratio of orbital to rotational angular momentum to
be $\alpha \sim 30$. In Fig.~4 of that paper, for this value of $\alpha$ the
time scale for circularization is much longer than the time for alignment.
Therefore, if $e$ = 0, which is the case of our system, no misalignment is
expected. Another explanation could be the loss of angular momentum due to
mass transfer. However, the system components are both nearly spherical and
far from filling the Roche lobe. The most plausible explanation seems to be
the radius expansion of the primary component related to its stage of
evolution. This star is passing through a region on the H-R diagram of
roughly constant luminosity, decreasing effective temperature, and
increasing radius. According to the grid of stellar models with rotation of
\citet{Ekstrometal2012}, a star of about 2~\Msun\ that has just evolved off
the main sequence will pass through a phase of decreasing equatorial
velocity before reaching the base of the giant branch (see their Fig.~9).
Though these models were computed for single stars, and not specifically for
a star with the same parameters as the primary component of our system, they
give us an indication that a decrease in the rotation rate is possibly
taking place.

The division of the light curve into 8 segments of about 20~days each was
necessary to identify the bona fide pulsation frequencies. A shift of the
frequency phases with time seems to disturb the identification of
frequencies in the whole time series. We performed several tests in order to
check the existence of phase shifts in the pulsation frequencies and to
identify any periodical variation. However, even if phase shifts are indeed
present, no clear periodical variation was found. We are not able to explain
the origin of this variation, though we think that it is likely intrinsic to
the star. The data reduction process and an instrument related effect could
be the cause, but we have applied the same method to several other light
curves of both CoRot and Kepler systems, and we did not see the same
behavior before.

We derived our results based on the first of the eight segments. We tested
the others by proceeding with the prewhitening steps, which led to new
solutions for the binary models, and to the identification of pulsation
frequencies in each segment. A comparison of the fitted parameters shows
very good agreement and low dispersion among the segments. The mean values
are:
$\langle$\teffs$\rangle$ = 6162 $\pm$ 13~K,
$\langle{i}\rangle$ = 81.66 $\pm$ 0.15~\degr,
$\langle{\Omega_1}\rangle$ = 4.27 $\pm$ 0.02,
$\langle{\Omega_2}\rangle$ = 7.89 $\pm$ 0.05,
$\langle{\beta_1}\rangle$ = 0.52 $\pm$ 0.02,
$\langle{A_1}\rangle$ = 0.77 $\pm$ 0.11,
$\langle{A_2}\rangle$ = 0.06 $\pm$ 0.08, and
$\langle{B}\rangle$ = 1.48 $\pm$ 0.04.
The dispersions are all smaller than the estimated uncertainties, and the
values agree with those in Table~\ref{par_tab}. Regarding the analysis of
the pulsation frequencies, the peaks with higher amplitudes in
Fig.~\ref{freq_spec} ($> 1\times 10^{-3}$) were normally identified in all
the eight segments of light curve, in particular the four genuine p-modes
($F_1$, $F_2$, $F_3$, and $F_5$) listed in Table~\ref{freq}. Small shifts in
amplitude, probably related to the phase shifts, were also observed among
the segments.

We believe it would be useful to have more spectra collected during the
eclipses in order to allow the modeling the Rossiter-McLaughlin effect.
This would confirm whether the spin-orbit axes are indeed aligned, as we
suspect, and reinforce the possibility of radius expansion of the primary
star due to its stage of evolution. The gathering of more spectra, with
higher S/N, would also improve the precision achieved in the spectroscopic
analysis, yielding to an accurate abundance determination of elements other
than iron. We also believe that the behavior of the amplitude and phase
variations is still not well understood. The development of more robust
programs and methods for the analysis of time series is required to properly
deal with this kind of data, in which a large number of observation points
and pulsation frequencies are present.

\begin{acknowledgements}
We thank Josefina Montalb\'an and Marc-Antoine Dupret for the non-adiabatic
calculation of the excited pulsation frequencies.
We are also very grateful for the referee report to this work, which has
contributed a lot to largely improve our manuscript.
This research has made use of the ExoDat Database, operated at LAM-OAMP,
Marseille, France, on behalf of the CoRoT/Exoplanet program, and
was accomplished with the help of the VO-KOREL cloud service, developed at
the Astronomical Institute of the Academy of Sciences of the Czech Republic
in the framework of the Czech Virtual Observatory (CZVO) by P. Skoda and
L. Mrkva using the Fourier disentangling code KOREL by P. Hadrava.
We acknowledge the generous financial support by the Istituto Nazionale di
Astrofisica (INAF) under Decree No. 28/2011 {\it Analysis and Interpretation
of CoRoT and Kepler data of single and binary stars of asteroseismological
interest}.
D.G. has received funding from the European Union Seventh Framework
Programme (FP7/2007-2013) under grant agreement n. 267251 (AstroFit).
A.P.H. acknowledges the support of DLR grant 50 OW 0204.
D.G. thanks John Kuehne and David Doss from McDonald Observatory, and Ivo
Saviane from ESO for their excellent support during the observations.
\end{acknowledgements}

\bibliographystyle{aa}
\bibliography{daSilvaetal2013}

\begin{thebibliography}{}

\bibitem[Asplund et al.(2005)]{Asplundetal2005}
   Asplund, M., Grevesse, N., \& Sauval, A.J. 2005, in Cosmic Abundance as
   Records of Stellar Evolution and Nucleosynthesis, eds. T.G. Barnes, III,
   \& F.N. Bash, ASP Conf. Ser., 336, 25
\bibitem[Auvergne et al.(2009)]{Auvergneetal2009}
   Auvergne, M., Bodin, P., Boisnard, L. et al. 2009, A\&A, 506, 401
\bibitem[Baglin et al.(2006)]{Baglinetal2006}
   Baglin, A., Auvergne, M., Boisnard, L., et al. 2006, 36th COSPAR
   Scientific Assembly, 36, 3749
\bibitem[Blackwell \& Shallis(1979)]{BlackwellShallis1979}
   Blackwell, D.E. \& Shallis, M.J. 1979, MNRAS, 186, 673
\bibitem[Bloemen et al.(2011)]{Bloemenetal2011}
   Bloemen, S., Marsh, T.R., \O stensen, R.H., et al. 2011, MNRAS, 410, 1787
\bibitem[Buzasi et al.(2005)]{Buzasietal2005}
   Buzasi, D.L., Bruntt, H., Bedding, T.R., et al. 2005, ApJ, 619, 1072
\bibitem[Claret(1999)]{Claret1999}
   Claret, A. 1999, ASP Conf. Ser., 173, 277
\bibitem[Charbonneau(2002)]{Charbonneau2002}
   Charbonneau, P. 2002, Pikaia genetic algorithms, version 1.2
\bibitem[Davidge \& Milone(1984)]{DavidgeMilone1984}
   Davidge, T.J., \& Milone, E.F. 1984, ApJS, 55, 571
\bibitem[Deleuil et al.(2009)]{Deleuiletal2009}
   Deleuil, M., Meunier, J.C., Moutou, C., et al. 2009, AJ, 138, 649
\bibitem[Dupret et al.(2005)]{Dupretetal2005}
   Dupret, M.-A., Grigahc\`ene, A., Garrido, R., et al. 2005, A\&A, 361, 476
\bibitem[Ekstr\"om et al.(2012)]{Ekstrometal2012}
   Ekstr\"om, S., Georgy, C., Eggenberger, P., et al. 2012, A\&A, 537, 146
\bibitem[Faigler \& Mazeh(2011)]{FaiglerMazeh2011}
   Faigler, S., \& Mazeh, T. 2011, MNRAS, 415, 3921
\bibitem[Ford(2005)]{Ford2005}
   Ford, E.B. 2005, AJ, 129, 1706
\bibitem[Gelman et al.(2003)]{Gelmanetal2003}
   Gelman, A., Carlin, J.B., Stern, H.S., \& Rubin, D.B. 2003, in Bayesian
   Data Analysis, London: Chapman \& Hall
\bibitem[Gray(1976)]{Gray1976}
   Gray, D.F. 1976, in The observation and analysis of stellar photospheres,
   Cambridge University Press, 2nd ed.
\bibitem[Gray \& Corbally(1994)]{GrayCorbally1994}
   Gray, R.O., \& Corbally, C.J. 1994, AJ, 107, 742
\bibitem[Gruberbauer(2008)]{Gruberbauer2008}
   Gruberbauer, M. 2008, n2XX - CoRoT n2 data eXplorer/eXtractor - Manual
   V1.2a
\bibitem[Hadrava(2004)]{Hadrava2004}
   Hadrava, P. 2004, Publ. Astron. Inst. Acad. Sci. Czech Rep., 92, 15
\bibitem[Hadrava(2009)]{Hadrava2009}
   Hadrava, P. 2009, ArXiv:0909.0172 [astro-ph.SR]
\bibitem[Hensberge et al.(2008)]{Hensbergeetal2008}
   Hensberge, H., Iliji\'c, S., \& Torres, K.B.V. 2008, A\&A, 482, 1031
\bibitem[Hut(1981)]{Hut1981}
   Hut, P. 1981, A\&A, 99, 126
\bibitem[Kaufer et al.(1999)]{Kaufer1999}
   Kaufer, A., Stahl, O., Tubbesing, S., et al. 1999, The Messenger, 95, 8
\bibitem[Kurucz(1993)]{Kurucz1993}
   Kurucz, R. 1993, CD-ROM No.~13, ATLAS\,9 Stellar Atmosphere Programs and
   2~\kms\ Grid (Cambridge, Mass.: Smithsonian Astrophysical Observatory)
\bibitem[Lenz \& Breger(2005)]{LenzBreger2005}
   Lenz, P., \& Breger, M. 2005, Comm. Asteroseismol., 146, 53
\bibitem[Loeb \& Gaudi(2003)]{LoebGaudi2003}
   Loeb, A., \& Gaudi, B.S. 2003, ApJ, 588, L117
\bibitem[Lucy(1967)]{Lucy1967}
   Lucy, L.B. 1967, Zeitschrift f\"ur Astrophysik, 65, 89
\bibitem[Maceroni et al.(2009)]{Maceronietal2009}
   Maceroni, C., Montalb\'an, J., Michel, E., et al. 2009, A\&A, 508, 1375
\bibitem[Maceroni et al.(2014)]{Maceronietal2014}
   Maceroni, C., Lehmann, H., da Silva, R., et al. 2014, accepted for
   publication in A\&A
\bibitem[Mazeh \& Faigler(2010)]{MazehFaigler2010}
   Mazeh, T., \& Faigler, S. 2010, A\&A, 521, L59
\bibitem[McCarthy et al.(1993)]{McCarthyetal1993}
   McCarthy, J.K., Sandiford, B.A., Boyd, D., \& Booth, J. 1993, PASP, 105,
   881
\bibitem[Milone(1968)]{Milone1968}
   Milone, E.E. 1968, AJ, 73, 708
\bibitem[Montgomery \& O'Donoghue(1999)]{MontgomeryODonoghue1999}
   Montgomery, M.H., \& O'Donoghue, D. 1999, DSSN, 13, 28
\bibitem[Nidever et al.(2002)]{Nideveretal2002}
   Nidever, D.L., Marcy, G.W., Butler, R.P., Fischer, D.A., \& Vogt, S.S.
   2002, ApJS, 141, 503
\bibitem[O'Connell(1951)]{OConnell1951}
   O'Connell, D.J.K. 1951, Publ. Riverview College Obs., 2, 85
\bibitem[Pr$\check{\rm s}$a \& Zwitter(2005)]{PrsaZwitter2005}
   Pr$\check{\rm s}$a, A., \& Zwitter, T. 2005, ApJ, 628, 426
\bibitem[Rodr\'iguez \& Breger(2001)]{RodriguezBreger2001}
   Rodr\'iguez, E., \& Breger, M. 2001, A\&A, 366, 178
\bibitem[Simon \& Sturm(1994)]{SimonSturm1994}
   Simon, K.P., \& Sturm, E. 1994, A\&A, 281, 286
\bibitem[Stellingwerf(1979)]{Stellingwerf1979}
   Stellingwerf, R.F. 1979, ApJ, 227, 935
\bibitem[Udry et al.(1999)]{Udryetal1999}
   Udry, S., Mayor, M., \& Queloz, D. 1999, ASP Conf. Ser., 185, 367
\bibitem[Uytterhoeven et al.(2011)]{Uytterhoevenetal2011}
   Uytterhoeven, K., Moya, A., Grigahc\`ene, A., et al. 2011, A\&A, 534, 125
\bibitem[Wilson \& Devinney(1971)]{WilsonDevinney1971}
   Wilson, R.E., \& Devinney, E.J. 1971, ApJ, 166, 605
\bibitem[Yi et al.(2003)]{Yietal2003}
   Yi, S.K., Kim, Y.-C., \& Demarque, P. 2003, \apjs, 144, 259
\bibitem[Zucker et al.(2007)]{Zuckeretal2007}
   Zucker, S., Mazeh, T., \& Alexander, T. 2007, \apj, 670, 1326
\end{thebibliography}

\end{document}